\documentclass[twocolumn,aps,pra,longbibliography,showpacs,superscriptaddress,nofootinbib,10pt]{revtex4-1}
\usepackage[latin1]{inputenc}
\usepackage{graphicx,dcolumn,bm}
\usepackage{subfigure}
\usepackage{color}
\usepackage[colorlinks,hyperindex]{hyperref}
\usepackage{amsfonts}
\usepackage{amssymb}
\usepackage{amsmath}
\usepackage{latexsym}
\usepackage{times,txfonts}
\usepackage{epstopdf}

\usepackage{multibib}

\usepackage{mathrsfs}
\usepackage{arydshln}
\usepackage[sans]{dsfont}

\usepackage{epsfig}

\definecolor{Blue}{rgb}{0.0,0.0,1}
\definecolor{Red}{rgb}{1,0.0,0.0}
\definecolor{Green}{rgb}{0,0.5,0.0}

\usepackage{tikz}
\usetikzlibrary{decorations.pathmorphing}
\usetikzlibrary{arrows}
\usetikzlibrary{intersections,shapes.arrows}

\usetikzlibrary{calc}
\usetikzlibrary{quotes,angles}
\usepackage{nicefrac}

\usepackage{pgfplots}
\usepgfplotslibrary{fillbetween}
\pgfplotsset{compat=1.10}

\usepgfplotslibrary{groupplots} 
\usepgfplotslibrary[groupplots] 
\usetikzlibrary{pgfplots.groupplots} 
\usetikzlibrary[pgfplots.groupplots] 
\usepgfplotslibrary{statistics}
\usepackage{pgfplotstable}
\tikzset{jumpdot/.style={mark=*,solid},excl/.append style={jumpdot,fill=white},incl/.append style={jumpdot,fill=black}}

\begin{document}

\title[Coherence orders, decoherence, and quantum metrology]
{Coherence orders, decoherence, and quantum metrology}

\author{Diego P. Pires}
\affiliation{International Institute of Physics, Universidade Federal do Rio Grande do Norte, Campus Universitario, Lagoa Nova, Natal-RN 59078-970, Brazil}
\author{Isabela A. Silva}
\affiliation{Instituto de Física de São Carlos, Universidade de São Paulo, CP 369, 13560-970, São Carlos, SP, Brazil}
\author{Eduardo R. deAzevedo}
\affiliation{Instituto de Física de São Carlos, Universidade de São Paulo, CP 369, 13560-970, São Carlos, SP, Brazil}
\author{Diogo O. Soares-Pinto}
\affiliation{Instituto de Física de São Carlos, Universidade de São Paulo, CP 369, 13560-970, São Carlos, SP, Brazil}
\author{Jefferson G. Filgueiras}
\affiliation{Instituto de Física de São Carlos, Universidade de São Paulo, CP 369, 13560-970, São Carlos, SP, Brazil}

\begin{abstract}
Since the dawn of quantum theory, coherence has been attributed as a key to understand the weirdness of fundamental concepts such as, e.g., the wave-particle duality and the Stern-Gerlach experiment. Recently, based on a resource theory approach, the notion of quantum coherence was revisited and a plethora of coherence quantifiers was proposed. In this work, we address such issue employing the language of coherence orders developed by the NMR community. This allowed us to investigate the role played by different subspaces of the Hilbert-Schmidt space into physical processes and quantum protocols. We found some links between decoherence and each coherence order. Moreover, we propose a sufficient and straightforward method to testify the usefulness of a given state for quantum enhanced phase estimation, relying on a minimal set of elements belonging to the density matrix.
\end{abstract}


\maketitle

\section{Introduction}

Quantum coherence is one of the most essential features of quantum theory. For example, a projective measurement over a pure state gets a deterministic outcome only if the state has no coherence with respect to the basis defined by the measurement projector~\cite{Jauch}. Moreover, coherence plays a key role in quantum information science, quantum thermodynamics, condensed matter physics and life sciences~\cite{2013_NatureCommun_4_2059,2013_arxiv_1302.2811,2013_arxiv_1308.1245,PhysRevLett.115.210403,PhysRevX.5.021001,2015_Lostaglio_Jennings_Rudolph_NatureCommun_4_6383,PNAS_2015_Branda_Horodecki_17032015,Acin2017,Gour2015,A_Rey_nature_13_781,A_Rey_arxiv_1706.01616,2013_Nature_9_10,Huelga_Plenio_10.1080_00405000.2013.829687,2014_Nature_10_621}.

There are several recent studies reporting the role of coherence in the framework of resource theories~\cite{PhysRevLett.116.120404,PhysRevA.95.062327,PhysRevLett.117.030401}, i.e., a successful theoretical approach to characterize entanglement~\cite{PhysRevLett.116.240405} and quantum thermodynamics~\cite{PhysRevLett.111.250404,1751-8121-49-14-143001,1367-2630-18-2-023045,PhysRevX.6.041017}. The resource theory for quantum coherence is similar to the theory of asymmetry~\cite{PRASpekkens1,PRASpekkens2}. A quantum state may or may not be invariant under the action of a given symmetry group. Thus, the degree of asymmetry quantifies how much the symmetry is broken by the quantum state. In this view, coherence is interpreted as the asymmetry respective to the group of translations ge\-ne\-ra\-ted by an observable, e.g., the Hamiltonian or one of the components of the angular momentum vector \cite{MarvianThesis}.

A wide variety of quantum coherence quantifiers was proposed in the literature during the last years~\cite{BCP2014,1751-8121-50-4-045301}. For example, such measures include both robustness and distillable coherence, convex roof quantifiers, coherence monotones and others~\cite{RevModPhys.89.041003}. In particular, the so-called distance-based coherence quantifiers provide a geometric viewpoint in the characterization of quantum coherence. In summary, such quantifiers are simply related to the total amount of coherence exhibited by the quantum state. However, this information alone is not enough in some physical situations. For example, concerning phase estimation protocols, the maximally coherent state, $\frac{1}{2^{N/2}}{(|{0}\rangle + |{1}\rangle)^{\otimes N}}$, offers smaller precision than a Greenberger-Horne-Zeilinger (GHZ) state. Interestingly, the latter exhibits less coherence than the former according to distance-based measures such as the $\ell_{1}$ norm and the relative entropy of coherence~\cite{BCP2014}. As pointed out by Marvian and Spekkens~\cite{MS2016}, this happens because such quantifiers deal only with speakable coherence. If the way to encode coherence is not relevant for a certain task, it denotes speakable coherence. Otherwise, like in the phase estimation example above, unspeakable coherence depends on how the information is encoded in the quantum state. 

Here we will focus on unspeakable coherence by ex\-ploi\-ting the concept of coherence order, i.e., an operational language developed by the nuclear magnetic resonance (NMR) community over the past $60$ years \cite{Munowitz,Keeler,Ernst2}. For example, the idea of coherence order finds application in multidimensional NMR spectroscopy in which the two or more frequency dimensions (the sets of frequencies being probed) are correlated through coherence orders. This kind of multidimensional spectrum provides information on the 3D structure of large molecules in biological and polymer samples, as well as in inorganic glasses~\cite{Ernst1,SchmidtRohr,Oigreslignina}. In many applications, the se\-cond coherence order plays an important role since it indicates only spin pairs interacting through the residual dipolar coupling, which allows one to estimate the average distance between coupled spins in an amorphous solid~\cite{elas1,elas2,elas3}. Furthermore, when simulating localization effects induced by decoherence through a spin counting experiment~\cite{Gonzalo1,Gonzalo2,Gonzalo3,Gonzalo4}, one can verify the number of correlated spins measuring the distribution of the signal among all coherence orders.

Employing this language, here we define quantifiers for each coherence order and discuss how they behave in open quantum systems subject to dephasing, highlighting their relation to decoherence-free subspaces. Moreover, we propose a simple and straightforward criterion to show the usefulness of a state for quantum enhanced metrology, built upon the concepts of multiple-quantum intensity (MQI) and the squared speed proposed in Refs.~\cite{PhysRevA.96.042327,A_Rey_arxiv_1706.01616}, respectively. This criterion has the advantage of using a minimal set of measurements in comparison to any other figure of merit found in li\-te\-ra\-tu\-re and holds for any number of qubits, even in the context of mixed-state quantum me\-tro\-lo\-gy~\cite{ModiPRX}.

\section{Measures of coherence}
\label{sec:measurescoherence002}

The quantification of coherence was initially proposed by \AA{}berg~\cite{Aberg1,Aberg2} and recently updated by Baumgratz {\it et al.}~\cite{BCP2014} according to an axiomatic framework inspired on resource theories. Both approaches deal with speakable coherence and are based on a small set of properties that any coherence quantifier $C$ must fulfill. To enunciate such properties, it is essential to establish the concepts of incoherent states and operations. Moreover, any discussion about quantum coherence requires the choice of a preferred basis of states. Thus, given a $d$-dimensional Hilbert space $\mathcal{H}$, let us fix the reference orthonormal basis ${\{|j\rangle\}_{j = 1,\ldots,d}}$. Incoherent states (IS) are those with all the off-diagonal elements equal to zero in this basis. The set of incoherent states $\mathcal{I}$ is a subset of the space of quantum states. On the other hand, an incoherent operation (IO) $\Lambda(\bullet)$ is a completely positive trace-preserving map that does not create coherence when acting over the set of incoherent states $\mathcal{I}$, i.e., $\Lambda[\mathcal{I}] \subseteq \mathcal{I}$. There are four rules to be fulfilled by $C$~\cite{BCP2014,RevModPhys.89.041003}: (i) \emph{non-negativity}: $C(\rho) \geq 0$, with the equality if and only if $\rho$ is incoherent; (ii) \emph{monotonicity}: $C(\Lambda[\rho])\leq C(\rho)$, for any $\rho$ and any incoherent operation $\Lambda$; (iii) \emph{convexity}: ${\sum_j}\,{p_j}C({\rho_j}) \geq C({\sum_j}\,{p_j}\,{\rho_j})$, for $0 \leq {p_j} \leq 1$ and any states ${\rho_j}$; and (iv) \emph{strong monotonicity}: $C$ does not increase, on average, under selective incoherent operations: ${\sum_j}\,{q_j}\,C({\sigma_j}) \leq C(\rho)$, with $q_j = \text{Tr}\,[{K_j}\, \rho {K_j^{\dagger}}]$, post-measurement states ${\sigma_j} = {K_j}\, \rho {K_j^{\dagger}}/{q_j}$, and incoherent Kraus operators ${K_j}$.

Interestingly, there is an interplay between incoherent states and quantum operations in coherence theory and both sets of separable states and local operations and classical communication (LOCC) in entanglement theory. Since both sets of states and operations share some properties, e.g., convexity when dealing with states, then some entanglement quantifiers can be rephrased for the coherence scenario. This is the case of the relative entropy of coherence~\cite{BCP2014} and the robustness of coherence (or asymmetry)~\cite{AdessoRobust1,AdessoRobust2}, where the latter represents a straightforward adaptation of the generalized robustness of entanglement~\cite{Steiner}. When considering the resource theory scenario, the robustness of coherence (or entanglement) defines a measure with a clear operational meaning, namely the minimum mixing to make a state incoherent (or separable). Reference~\cite{RevModPhys.89.041003} provides a recent review on the subject of quantum coherence as a resource. Here we intend to present an alternative and operationally simpler formulation to characterize quantum coherence. Furthermore, we show that such a proposal is equivalent to the $\textsf{U}(1)$ resource theory of asymmetry.

The quantification of unspeakable coherence has been addressed recently~\cite{MS2016} based on the asymmetry relative to a group of translations. Let us consider the unitary representation of a group of translations generated by an observable $H$ describing any re\-le\-vant phy\-si\-cal quantity, e.g., energy or angular momentum. This representation is defined as
\begin{equation}
\label{1a}
{U_{H,x}} = {e^{-ixH}}, \quad x\in\mathbb{R} ~,
\end{equation}
and the action of such group on a state $\rho$ is given by
\begin{equation}
\label{1b}
{U_{H,x}}[\rho] = {e^{-ixH}}\rho\,{e^{ixH}} ~.
\end{equation}
In the approach of translationally covariant operations, the incoherent states are those invariant under the translations, i.e, states commuting with the generator $H$. Therefore, the definition of coherence is relative to the eigenspaces of the obser\-va\-ble $H$. The incoherent operations are those which are covariant with respect to the symmetry group, i.e., a quantum operation $\mathcal{E}(\bullet)$ is translationally covariant if 
\begin{equation}
\label{eq:TI000001}
{U_{H,x}}[\mathcal{E}\,(\rho)] = \mathcal{E}\,({U_{H,x}}[\rho]) ~, \quad \forall x\in\mathbb{R} ~,~ \forall \rho ~.
\end{equation}

The idea of coherence via translationally covariant operations allows us to separate the coherence of each invariant subspace of $H$ using the concept of modes of asymmetry~\cite{PRASpekkens1}. The modes of asymmetry are related to the projectors $P^{(m)}$ defined as
\begin{equation}
\label{2}
P^{(m)}(\rho) = {\lim_{{x_0}\rightarrow \infty}} \,\frac{1}{2{x_0}}\,{\int_{-x_0}^{x_0}}\,dx\, {e^{-i m x}}\, {U_{H,x}}[\rho] ~.
\end{equation}
The index $m$ is associated with the difference of the eigen\-va\-lues of $H$, i.e., the gaps in the spectrum of the Hamiltonian $H$. The set of projectors ${P^{(m)}}$, $\forall m$, specifies a complete orthogonal basis of the Hilbert-Schmidt space and also provides the decomposition of states, operations and measurements as linear combinations of ${P^{(m)}}$'s. The subspaces associated with different values of $m$ are termed modes of asymmetry~\cite{PRASpekkens1,PRASpekkens2}.

The quantifier of unspeakable coherence of a state $\rho$ is set as
\begin{equation}
\label{3}
{C_{m}}(\rho) = {\| {P^{(m)}}(\rho)\|_1} ~,
\end{equation}
where ${\|{X}\|_1} := \text{Tr}\,(\sqrt{X{X^{\dagger}}} \,)$ is the trace norm and $C$ is defined for each $m \ne 0$. We remark that such translationally covariant measures of coherence are not, in general, coherence measures according to the properties (i)--(iv). For a specific $m$, $C_{m}$ can increase under incoherent operations because these operations are able to move coherence from other modes to the mode-$m$ component~\cite{MS2016}. 

Opposite to the trace norm, the so-called Hilbert-Schmidt (HS) norm or Schatten $2$-norm, i.e., ${\|{X}\|_2} := \sqrt{\text{Tr}\,(X{X^{\dagger}})}$, does not characterize itself as a \textit{bona fide} coherence quantifier according to the properties ($i$)--($iv$) introduced in Ref.~\cite{BCP2014}. The HS norm does not define a proper speakable coherence measure since it violates the monotonicity condition, i.e., it can increases under in\-co\-he\-rent operations. Furthermore, Ref.~\cite{MS2016} shows that the HS norm does not define a valid $\textsf{U}(1)$-asymmetry monotone. If a more restricted set of incoherent operations is considered, the HS norm defines a quantifier of unspeakable coherence in the context of the so-called resource theory of genuine coherence~\cite{1751-8121-50-4-045301}.

As a final remark, it is important to discuss the main differences between both ``incoherent'' notions introduced in the last paragraphs. According to the framework presented by Baumgratz {\it et al.}~\cite{BCP2014}, an incoherent state refers to a given density matrix that does not have coherence with respect to a fixed preferred basis. On the other hand, Marvian and Spekkens address the concept of incoherent state by cha\-rac\-te\-ri\-zing its degree of asymmetry respective to the action of a given symmetry group generator.

\section{Coherence orders}

In several applications of quantum mechanics, a proper choice of basis can often simplify the calculations and provide a clear physical significance to the quantities being e\-va\-lua\-ted. Thus, for a system of qubits, we adopt the operator basis $\{{I_0}, {I_+}, {I_-}, {I_z}\}$, which is widely used in the NMR literature, with ${I_{\pm}} = (\sigma_x \pm i\sigma_y)/2$ being the ladder operators, ${I_z} = {\sigma_z}/2$ is the $z$ component of the angular momentum, and ${I_0} = \mathbb{I}/2$ is the identity matrix~\cite{Munowitz,Keeler,Ernst2}. We remark that this particular basis is meaningful because it serves as an eigenbasis for any interaction whose Hamiltonian depends on a linear combination of $I_z$ and $I_{j_k}\otimes I_{j_l}$, with ${j_k} = \{0,x,y,z\}$, and the subindex indicating the $k$th qubit. Focusing on a system of $N$-qubits, one can represent the density matrix $\rho$ as follows:
\begin{equation}
\label{4}
\rho = \sum_{{j_1},{j_2},\ldots,{j_N}}{a_{{j_1}{j_2}\ldots{j_N}}} \, \bigotimes_{l = 1}^{N} \, {I_{j_l}} ~,  
\end{equation}
with ${j_l} = \{0,+,-,z\}$, the subindex $l \in \{1,\ldots,N\}$ indicating the particle number, and ${a_{{j_1}{j_2}\ldots{j_N}}} = \text{Tr}\,(\,\rho \, {I_{{j_1}{j_2}\ldots {j_N}}})$ is a complex number.

A clear advantage of this basis appears when we apply a rotation about the $z$ axis, as ${e^{-i\theta {I_z}}}{I_{j_l}}{e^{i\theta {I_z}}} = {e^{-i({\delta_{+{j_l}}} - {\delta_{-{j_l}}})\theta}}{I_{j_l}}$. Therefore, under a rotation generated by the $z$ component of the total angular momentum, an arbitrary $\bigotimes_{l = 1}^{N} \, {I_{j_l}}$ behaves as
\begin{equation}
\label{zrot}
{e^{-i\theta Z}} \bigotimes_{l = 1}^{N} \, {I_{j_l}} \,{e^{i\theta Z}} = {e^{-im\theta}} \bigotimes_{l = 1}^{N} \, {I_{j_l}} ~,
\end{equation}
where
\begin{equation}
Z := \sum_{l=1}^{N}\,{\mathbb{I}^{\, \otimes{l - 1}}}\otimes{I_z}\otimes{\mathbb{I}^{\, \otimes{N - l}}} ~,
\end{equation} 
and
\begin{equation}
m := {n_+} - {n_-}
\end{equation} 
defines the coherence order of $\bigotimes_{l = 1}^{N} \, {I_{j_l}}$ relative to the eigenbasis of $Z$. The number of times that ${I_{s}}$ appears in the decomposition of $\bigotimes_{l = 1}^{N} \, {I_{j_l}}$ is given by ${n_s} = {\sum_{l = 1}^{N}{\delta_{s,{j_l}}}}$ ($s = \{0,+,-,z\}$). Moreover, the set of $n_s$'s fulfills the constraint ${n_0} + {n_+} + {n_-} + {n_z} = N$. Because the density matrix in Eq.~\eqref{4} is Hermitian, each coherence order always occurs in pairs $\pm m$. This pro\-per\-ty allows us to rewrite $\rho$ in a decomposition of subspaces related to each particular coherence order as~\cite{2009_Int_J_Quantum_Inform_7_125_Paris}
\begin{equation}
\label{smartdm}
\rho = {\sum_{m = -N}^{N}\, {\rho_m}} ~,
\end{equation}
with
\begin{equation}
{\rho_m} := {\sum_{{n_+} - {n_-} = m}}\, {a_{{j_1}{j_2}\ldots{j_N}}} \, {\bigotimes_{l = 1}^{N}} \, {I_{j_l}}
\end{equation}
being the projection of $\rho$ into the subspace spanned by $\bigotimes_{l = 1}^{N} \, {I_{j_l}}$ related to the $m$th coherence order. Notice that ${\rho_m^{\dagger}} = {\rho_{-m}}$ $\forall m \in \{-N,\ldots,N\}$. Another important property is that all $\rho_m$ are orthogonal with respect to the Hilbert-Schmidt inner product, i.e., $\text{Tr}({\rho_m^{\dagger}}\,{\rho_n}) = \text{Tr}({\rho_{-m}}\,{\rho_n}) \propto{\delta_{mn}}$. Furthermore, for nonzero ${n_+}$ and/or ${n_-}$, each $\rho_m$ has $N_m$ elements, given by
\begin{equation}
\label{mcohnumber}
{N_m} = \frac{(2N)!}{{(N-m)!}{(N+m)!}} ~,
\end{equation}
which holds for $m\ne 0$~\cite{Munowitz}.

\begin{figure}[t]
\centering
\includegraphics[width = 1.0\columnwidth]{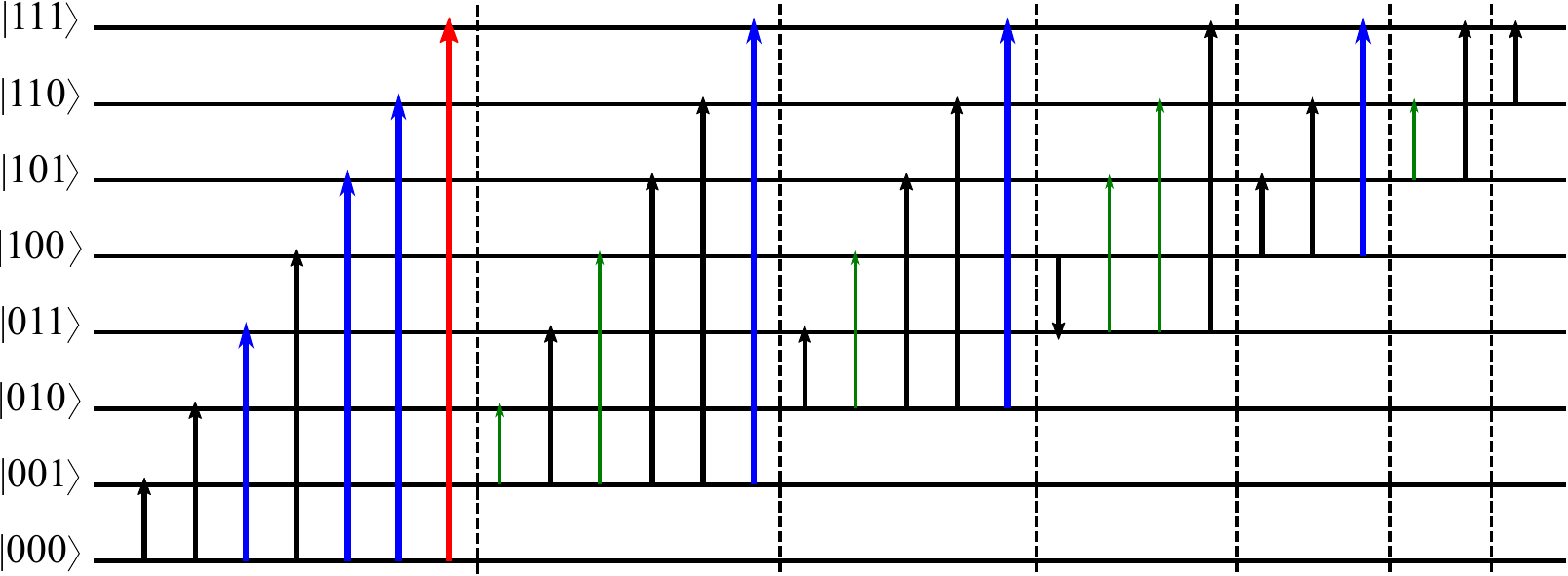}
\caption{(Color online) Depiction of all possible transitions of a three-qubit system and their relation to each coherence order. The third coherence order in red (thicker), second in blue, first in black, and zeroth in green (thinnest). The arrows indicate the direction of a positive order. The dashed lines separate the transitions associated with each line of Fig.~\ref{fig:orders}(c).}
\label{fig:levels}
\end{figure}

\begin{figure*}[t]
\subfigure[ref1][~]{
\begin{tikzpicture}[scale=0.8]
\node[black] (a) at (0,1) {{$0$}};
\node[black] (b) at (1,1) {{$+1$}};
\node[black] (c) at (0,0) {{$-1$}};
\node[black] (d) at (1,0) {{$0$}};
\draw[black,very thick] (-0.5,-0.5) -- (-0.5,1.5);
\draw[black,very thick] (-0.5,1.5) -- (-0.35,1.5);
\draw[black,very thick] (-0.5,-0.5) -- (-0.35,-0.5);
\draw[black,very thick] (1.5,-0.5) -- (1.5,1.5);
\draw[black,very thick] (1.35,1.5) -- (1.5,1.5);
\draw[black,very thick] (1.35,-0.5) -- (1.5,-0.5);
\node[black,very thick] (ia) at (0.5,2) {{$N = 1$}};
\end{tikzpicture}}
\quad\quad
\subfigure[ref2][~]
{\begin{tikzpicture}[scale=0.8]
\draw[black,very thick] (-0.5,-0.5) -- (-0.5,3.5);
\draw[black,very thick] (-0.5,3.5) -- (-0.35,3.5);
\draw[black,very thick] (-0.5,-0.5) -- (-0.35,-0.5);
\draw[black,very thick] (3.5,-0.5) -- (3.5,3.5);
\draw[black,very thick] (3.35,3.5) -- (3.5,3.5);
\draw[black,very thick] (3.35,-0.5) -- (3.5,-0.5);
\draw[black,dashed,very thick] (1.5,-0.5) -- (1.5,3.5);
\draw[black,dashed,very thick] (-0.4,1.5) -- (3.4,1.5);
\node[black,very thick] (a) at (0,3) {{$0$}};
\node[black] (b) at (1,3) {{$+1$}};
\node[black] (c) at (2,3) {{$+1$}};
\node[black] (d) at (3,3) {{$+2$}};
\node[black] (e) at (0,2) {{$-1$}};
\node[black] (f) at (1,2) {{$0$}};
\node[black] (g) at (2,2) {{$0$}};
\node[black] (h) at (3,2) {{$+1$}};
\node[black] (i) at (0,1) {{$-1$}};
\node[black] (j) at (1,1) {{$0$}};
\node[black] (k) at (2,1) {{$0$}};
\node[black] (l) at (3,1) {{$+1$}};
\node[black] (m) at (0,0) {{$-2$}};
\node[black] (n) at (1,0) {{$-1$}};
\node[black] (o) at (2,0) {{$-1$}};
\node[black] (p) at (3,0) {{$0$}};
\node[black,very thick] (ia) at (1.5,4) {{$N = 2$}};
\end{tikzpicture}}
\quad\quad
\subfigure[ref3][~]{
\centering
\begin{tikzpicture}[scale=0.7]
\draw[black,very thick] (-0.5,-0.5) -- (-0.5,7.5);
\draw[black,very thick] (-0.5,7.5) -- (-0.35,7.5);
\draw[black,very thick] (-0.5,-0.5) -- (-0.35,-0.5);
\draw[black,very thick] (7.5,-0.5) -- (7.5,7.5);
\draw[black,very thick] (7.35,7.5) -- (7.5,7.5);
\draw[black,very thick] (7.35,-0.5) -- (7.5,-0.5);
\draw[black,dashed,very thick] (3.5,-0.5) -- (3.5,7.5);
\draw[black,dashed,very thick] (-0.4,3.5) -- (7.4,3.5);
\node[black,very thick] (aa) at (0,7) {{$0$}};
\node[black,very thick] (ba) at (1,7) {{$+1$}};
\node[black,very thick] (ca) at (2,7) {{$+1$}};
\node[black,very thick] (da) at (3,7) {{$+2$}};
\node[black,very thick] (ea) at (4,7) {{$+1$}};
\node[black,very thick] (fa) at (5,7) {{$+2$}};
\node[black,very thick] (ga) at (6,7) {{$+2$}};
\node[black,very thick] (ha) at (7,7) {{$+3$}};
\node[black,very thick] (ab) at (0,6) {{$-1$}};
\node[black,very thick] (bb) at (1,6) {{$0$}};
\node[black,very thick] (cb) at (2,6) {{$0$}};
\node[black,very thick] (db) at (3,6) {{$+1$}};
\node[black,very thick] (eb) at (4,6) {{$0$}};
\node[black,very thick] (fb) at (5,6) {{$+1$}};
\node[black,very thick] (gb) at (6,6) {{$+1$}};
\node[black,very thick] (hb) at (7,6) {{$+2$}};
\node[black,very thick] (ac) at (0,5) {{$-1$}};
\node[black,very thick] (bc) at (1,5) {{$0$}};
\node[black,very thick] (cc) at (2,5) {{$0$}};
\node[black,very thick] (dc) at (3,5) {{$+1$}};
\node[black,very thick] (ec) at (4,5) {{$0$}};
\node[black,very thick] (fc) at (5,5) {{$+1$}};
\node[black,very thick] (gc) at (6,5) {{$+1$}};
\node[black,very thick] (hc) at (7,5) {{$+2$}};
\node[black,very thick] (ad) at (0,4) {{$-2$}};
\node[black,very thick] (bd) at (1,4) {{$-1$}};
\node[black,very thick] (cd) at (2,4) {{$-1$}};
\node[black,very thick] (dd) at (3,4) {{$0$}};
\node[black,very thick] (ed) at (4,4) {{$-1$}};
\node[black,very thick] (fd) at (5,4) {{$0$}};
\node[black,very thick] (gd) at (6,4) {{$0$}};
\node[black,very thick] (hd) at (7,4) {{$+1$}};
\node[black,very thick] (ae) at (0,3) {{$-1$}};
\node[black,very thick] (be) at (1,3) {{$0$}};
\node[black,very thick] (ce) at (2,3) {{$0$}};
\node[black,very thick] (de) at (3,3) {{$+1$}};
\node[black,very thick] (ee) at (4,3) {{$0$}};
\node[black,very thick] (fe) at (5,3) {{$+1$}};
\node[black,very thick] (ge) at (6,3) {{$+1$}};
\node[black,very thick] (ge) at (7,3) {{$+2$}};
\node[black,very thick] (he) at (7,5) {{$+2$}};
\node[black,very thick] (af) at (0,2) {{$-2$}};
\node[black,very thick] (bf) at (1,2) {{$-1$}};
\node[black,very thick] (cf) at (2,2) {{$-1$}};
\node[black,very thick] (df) at (3,2) {{$0$}};
\node[black,very thick] (ef) at (4,2) {{$-1$}};
\node[black,very thick] (ff) at (5,2) {{$0$}};
\node[black,very thick] (gf) at (6,2) {{$0$}};
\node[black,very thick] (hf) at (7,2) {{$+1$}};
\node[black,very thick] (ag) at (0,1) {{$-2$}};
\node[black,very thick] (bg) at (1,1) {{$-1$}};
\node[black,very thick] (cg) at (2,1) {{$-1$}};
\node[black,very thick] (dg) at (3,1) {{$0$}};
\node[black,very thick] (eg) at (4,1) {{$-1$}};
\node[black,very thick] (fg) at (5,1) {{$0$}};
\node[black,very thick] (gg) at (6,1) {{$0$}};
\node[black,very thick] (hg) at (7,1) {{$+1$}};
\node[black,very thick] (ah) at (0,0) {{$-3$}};
\node[black,very thick] (bh) at (1,0) {{$-2$}};
\node[black,very thick] (ch) at (2,0) {{$-2$}};
\node[black,very thick] (dh) at (3,0) {{$-1$}};
\node[black,very thick] (eh) at (4,0) {{$-2$}};
\node[black,very thick] (fh) at (5,0) {{$-1$}};
\node[black,very thick] (gh) at (6,0) {{$-1$}};
\node[black,very thick] (hh) at (7,0) {{$0$}};
\node[black,very thick] (ia) at (3.5,8) {{$N = 3$}};
\end{tikzpicture}}
\caption{Coherence orders for (a)~$N = 1$, (b)~$N = 2$, and (c)~$N = 3$. Note the pattern as $N$ increases.}
\label{fig:orders}
\end{figure*}

Each $\rho_m$ is a projection on the subspace defined by the $m$-th coherence order. This is similar for modes of asymmetry, as they are related to a set of projectors ${P^{(m)}}(\rho)$ defined in Eq.~\eqref{2}. Using the relation $\rho = \sum_{m} \rho_m$ into Eq.~\eqref{2}, with $H = Z$, we get 
\begin{align}
\label{proof}
{P^{(m)}}(\rho) &=  {\sum_{m' = -N}^N}\, {\rho_m'} \, {\lim_{{x_0} \rightarrow \infty}}\, \frac{1}{2{x_0}}\,{\int_{- {x_0}}^{x_0}}\,dx \,{e^{-i(m + m')x}} \nonumber \\
&= {\sum_{m' = -N}^N}\, {\delta_{m,-m'}} \, {\rho_m'} = {\rho_{-m}} ~,
\end{align}
where we have recognized the definition of the Kronecker $\delta$ involving $m$ and $m'$. 
In summary, Eq.~\eqref{proof} unveils the equivalence between both approaches of coherence orders and $\textsf{U}(1)$ modes of asymmetry.

Each element of the density matrix is related to how information is encoded in the quantum system. On the one hand, populations are related to the probability to detect the system in one of its eigenstates. On the other hand, each coherence unveils the net interference between two different basis states. According to Eq.~\eqref{zrot}, different coherence orders oscillate independently and with distinct frequencies under global (or local) rotations. Since the order of a projector ${\bigotimes_{l = 1}^N}\,{I_{j_l}}$ is defined by the number of times the operators $I_+$ and $I_-$ appear on it, each order is a direct measure of correlation. For example, the second and third orders are measures of bipartite and tripartite correlations.

The coherence orders are related to the transitions between the levels of a quantum system. For instance, a $m$th coherence order is generated after the interaction of the quantum system with at least that $m$ quanta of radiation from an external field, generating a coherent superposition. Thus, as shown in Fig.~\ref{fig:levels}, the double quantum coherences are associated to transitions between levels separated by two quanta, while the maximum coherence order involves a superposition between the most energetic and the ground states of the system~\cite{Munowitz}. Interestingly, the zeroth order encompasses transitions with an equal number of absorptions and emissions.

The coherence orders spread along the density matrix following a simple recipe. Each line of the quantum state is related to the transitions between a fixed eigenstate and all the other ones. The first line of Fig.~\ref{fig:orders}(c) is connected to the first block of transitions in Fig.~\ref{fig:levels}, the second line to the second block, and so on. Moreover, due to the tensor product structure of the Hilbert-Schmidt space, the matrix of \ref{fig:orders}(a) serves as a building block for the $N$-qubit case, with sums among the elements of each block instead of products. This pattern is clear from Figs.~\ref{fig:orders}(a)--\ref{fig:orders}(c).

\section{${\ell_1}$-norm and dephasing of coherence orders}

It is possible to define quantifiers for each coherence order, relative to global rotations about the $z$ axis, adapting different coherence quantifiers defined in the literature. Using the $\ell_1$ norm, we define the amount of coherence of order $m$ stored in the state $\rho$ as
\begin{equation}
\label{quantdef}
{\mathcal{C}_{|m|}^{\ell_1}} = \frac{1}{2}\, {\sum_{\vert{n_+} - {n_-}\vert = m}}{\left| {a_{{j_1}{j_2}\ldots{j_N}}} \right|} ~,
\end{equation}
with $n_{+}\neq 0$. We point out one subtle difference between our definition and the quantifiers proposed in Ref.~\cite{MS2016}. The zeroth coherence order is not considered in the approach based on modes of asymmetry since it is invariant under global $z$ rotations, i.e., a state which has only this kind of coherence is regarded as incoherent one. This is due to the difference between the notion of incoherent operations and states applied here (rooted on the work by Baumgratz {\it et al.}~\cite{BCP2014}) and that based on asymmetry with respect to $Z$. However, since $P^{(m)}(\rho) = \rho_{-m}$, these approaches are equivalent. 

The quantifiers of Eq.~\eqref{quantdef} are not measures of coherence according to the properties (i)--(iv), as they do not satisfy the monotonicity property. An incoherent operation can increase or decrease the amount of coherence of a particular order. For example, in a two-qubit system, a second-order coherence can be transferred to zeroth order by a local $\pi$ rotation about the $x$ axis, as ${e^{-i\frac{\pi}{2}{\sigma_x}\otimes\,\mathbb{I}}} \, ({I_+}\otimes{I_+})\, {e^{i\frac{\pi}{2}{\sigma_x}\otimes\,\mathbb{I}}} \longrightarrow {I_-}\otimes{I_+}$. However, it is a monotone under $\textsf{U}(1)$-covariant operations~\cite{MS2016}.

As an example, let us discuss the behavior of coherence orders under dephasing. Particularly, we focus on processes due to Gaussian noise described by a zero mean and an homogeneous autocorrelation function $K(t,{t'}) = K(t - {t'})$ as follows~\cite{PARIS2014256,BENEDETTI20142495,S0219749915600035}:
\begin{equation}
\label{eq:h1q0010}
{[{B_j}(t)]_B} = 0 ~,\qquad
{[{B_j}(t){B_l}({t'})]_B} = {\delta_{jl}}K(t - {t'}) ~,
\end{equation}
where ${{[\bullet]}_B} := {\int} \mathcal{D}[B(t)] \, \mathcal{P}[B(t)] \, \bullet$ defines the average performed over all the possible realizations of the process $B(t)$, each one occurring with probability $\mathcal{P}[B(t)]$. Just to be clear, it follows that ${\int} \mathcal{D}[B(t)] \, \mathcal{P}[B(t)] = 1$. A Gaussian process is fully described by its second-order autocorrelation function $K$, with its characteristic function given by
\begin{equation}
\label{eq:h1q001}
{\left[\exp\left(\pm i\kappa {\int_0^t}ds\,B(s)\right)\right]_B} = \exp\left( {-{\kappa^2}\beta(t) }\right) ~,
\end{equation}
where we define 
\begin{equation}
\label{eq:h1q0000022222}
\beta(t) = \frac{1}{2}{\int_0^t}{\int_0^t} ds \, d{s'}K(s - {s'}) ~.
\end{equation}
\par\noindent Similar results, under different contexts, are found in the quantum information literature to describe the decoherence of quantum registers \cite{heina2002,Ischi2005} and in NMR to describe spin-spin relaxation \cite{Stoll1977,Ernst1} and long-lived states~\cite{Levitlong1,Levittlong2,Levitlong3}. Recently, Ref.~\cite{PhysRevLett.119.010403} proposed a general framework to investigate the amplitude of a stochastic noise in a fluctuating many-body Hamiltonian system.

\begin{figure}[h]
\begin{tabular}{ll}
\begin{tikzpicture}[scale=0.9,font=\normalsize]
\begin{axis}[
  title={Common environment},
    enlargelimits=false,
    very thick,
    minor y tick num=1,
    legend style={draw=none},
    ymin=0, ymax=1.04,
    xlabel={$t$ (s)},
    ylabel={Coherence Order},
    legend entries={${\widetilde{C}_0}$,${\widetilde{C}_1}$,${\widetilde{C}_2}$,${\widetilde{C}_3}$},
    legend style={at={(0.95,0.6)},anchor=east}
]
\addplot+[
    mark=diamond*,
    mark options = {brown},
     mark size = 2pt,
    mark repeat=1.2,
    mark phase=1,
    brown,
    very thick,
    domain=0:0.5,
    samples=50,
]
{1};
\addplot[
    red,
    very thick,
    dashed,
    domain=0:0.5,
    samples=100,
]
{exp(-0.5*(10*x - 1 + exp(-10*x) ))};
\addplot+[
   mark=star,
   mark size = 4pt,
     mark repeat=2,
    mark phase=1,
   blue,
   very thick,
   dashdotted,
   domain=0:0.5,
   samples=50,
]
{exp(-4*0.5*(10*x - 1 + exp(-10*x)))};
\addplot[
   black,
   very thick,
   dotted,
   domain=0:0.5,
   samples=100,
]
{exp(-9*0.5*(10*x - 1 + exp(-10*x) ))};
\node at (axis cs:0.45,0.9) {{\large{(a)}}};
\end{axis}
\end{tikzpicture}
\\
\begin{tikzpicture}[scale=0.9,font=\normalsize]
\begin{axis}[
  title={Independent environments},
    enlargelimits=false,
    very thick,
    minor y tick num=1,
    legend style={draw=none},
    ymin=0, ymax=1.04,
    xlabel={$t$ (s)},
    ylabel={Coherence Order},
    legend entries={${\widetilde{C}_0}$,${\widetilde{C}_1}$,${\widetilde{C}_2}$,${\widetilde{C}_3}$},
    legend style={at={(0.95,0.6)},anchor=east}
]
\addplot+[
    mark=diamond*,
    mark options = {brown},
     mark size = 2pt,
    mark repeat=1.2,
    mark phase=1,
    brown,
    very thick,
    domain=0:1,
    samples=50,
]
{0.756833*exp(-8.2*x - 0.82*exp(-10*x)) + 0.560676*exp(-5.2*x - 0.52*exp(-10*x)) + 0.468316*exp(-3.4*x - 0.34*exp(-10*x))};
\addplot[
    red,
    very thick,
    dashed,
    domain=0:1,
    samples=100,
]
{0.463273*exp(-8.4*x - 0.84*exp(-10*x)) + 0.439659*exp(-5*x - 0.5*exp(-10*x)) + 0.367234*exp(-3.2*x - 0.32*exp(-10*x)) + 0.272054*exp(-0.2*x - 0.02*exp(-10*x))};
\addplot+[
   mark=star,
   mark size = 4pt,
     mark repeat=2,
    mark phase=1,
   blue,
   very thick,
   dashdotted,
   domain=0:1,
   samples=50,
]
{0.756833*exp(-8.2*x - 0.82*exp(-10*x)) + 0.560676*exp(-5.2*x - 0.52*exp(-10*x)) + 0.468316*exp(-3.4*x - 0.34*exp(-10*x))};
\addplot[
   black,
   very thick,
   dotted,
   domain=0:1,
   samples=100,
]
{2.31637*exp(-8.4*x - 0.84*exp(-10*x))};
\node at (axis cs:0.9,0.9) {{\large{(b)}}};
\end{axis}
\end{tikzpicture}
\end{tabular}
\caption{(Color online) Open-system dynamics for each coherence order for a maximally coherent three-qubit state, $|{{+}{+}{+}}\rangle$, coupled to Ornstein-Uhlenbeck environments, in the slow correlation time regime. In particular, we set $\Gamma/({2{\gamma^2}}) \approx 1$, with $\Gamma = 100$~${(\text{rad}/\text{s})}^2$, $\tau = 1/\gamma = 0.1$~s, and $\lambda_l = \{1, 0.8, 0.2\}$. (a) Case of a common environment. All coherence orders decay with a single rate. The higher the order, the faster is the decay. The zeroth order remains invariant under dephasing. (b) Case of three independent environments.  Notice that both ${\widetilde{C}_{0}}$ and ${\widetilde{C}_{2}}$ coincide along the range $0 \leq t~(s) \leq 1$. There is a multicomponent decay for all coherence orders but the third one. Moreover, there is no direct connection between the specific coherence order and how fast or slow is the decay due to dephasing. In both cases (a) and (b) we plot the normalized coherence orders ${\{{\widetilde{C}_{m}}\}_{m = 0,1,2,3}}$, with ${\widetilde{C}_{m}} := {\mathcal{C}_{|m|}^{\ell_1}}(\rho(t))/{\mathcal{C}_{|m|}^{\ell_1}}(\rho(0))$.}
\label{fig:commonfigure02}
\end{figure}
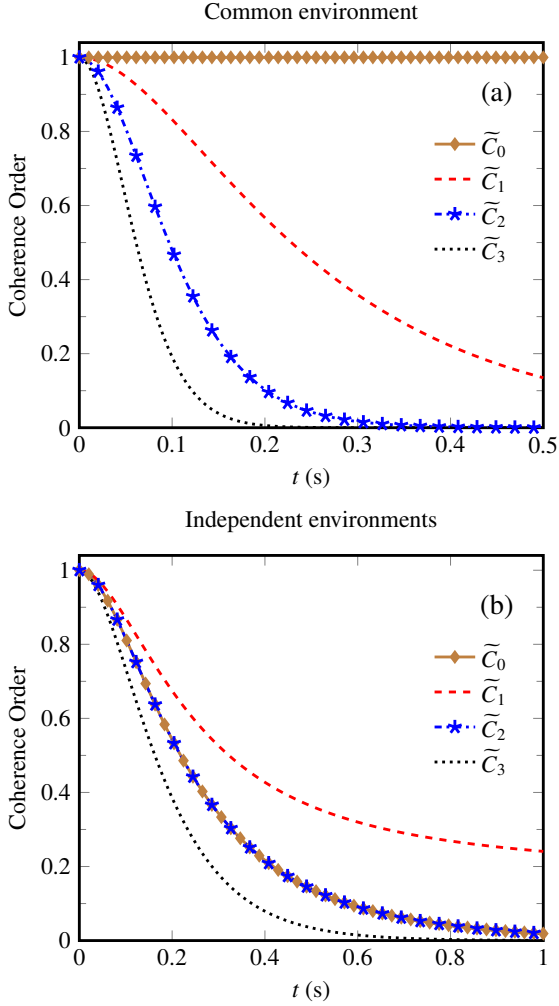

Let us consider the $N$-qubit state whose matrix representation is given by Eq.~\eqref{smartdm}. Considering the scenario in which each qubit is coupled to a single bath, the open-system dynamics is driven by the Hamiltonian $H(t) = {\sum_{l=1}^N}\,{\mathbb{I}^{\otimes{l - 1}}}\otimes{H_l}(t)\otimes{\mathbb{I}^{\otimes{N - l}}}$, with ${H_l}(t) = [{\omega_0} + \lambda_l{B_l}(t)]\,{I_z}$, where $\omega_0$ is the qubit energy splitting, $\lambda_l$ is the strength of the interaction, and ${B_l}(t)$ is an external classical stochastic field acting on each qubit. The evolution operator is given by
\begin{equation}
 \label{eq:Nqubitbelldiag0001002}
U(t) = \exp\left(-i{\int_0^t}ds H(s)\right) = {\bigotimes_{l=1}^N}\,{e^{-i{\varphi_l}(t){I_z}}} ~,
\end{equation}
where we define ${\varphi_l}(t) := {\omega_0}t + {\lambda_l}{h_l}(t)$ and ${h_l}(t) = {\int_0^t}ds \,{B_l}(s)$. Starting from Eqs.~\eqref{smartdm} and~\eqref{eq:Nqubitbelldiag0001002}, the evolved state ${\rho}(t) = {[U(t)\rho(0){U^{\dagger}}(t)]_B}$ can be written as follows
\begin{equation}
{\rho}(t) = {\sum_{{j_1},{j_2},\ldots,{j_N}}}{a_{{j_1}{j_2}\ldots{j_N}}} \, {\left[{\bigotimes_{l = 1}^N}\,
{e^{-i{\varphi_l}(t){I_z}}}{I_{j_l}}\,{e^{i{\varphi_l}(t){I_z}}}\right]_B} ~.
\end{equation}
Noticing that each element, under a $z$ rotation, behaves like ${e^{-i{\varphi_l}(t){I_z}}}{I_{j_l}}{e^{i{\varphi_l}(t){I_z}}} = {e^{-i({\delta_{+{j_l}}} - {\delta_{-{j_l}}}){\varphi_l}(t)}}{I_{j_l}}$, we have
\begin{equation}
\label{eq:generalstate001}
\rho(t) = {\sum_{{j_1},{j_2},\ldots,{j_N}}}{a_{{j_1}{j_2}\ldots{j_N}}}\, {\xi_{{j_1}{j_2}\ldots{j_N}}} \, \bigotimes_{l=1}^{N}{I_{j_l}} ~,
\end{equation}
with
\begin{equation}
\label{eq:auxfunction0001}
{\xi_{{j_1}{j_2}\ldots{j_N}}} = {\left[{\prod_{l = 1}^N}{e^{-i({\delta_{+{j_l}}} - {\delta_{-{j_l}}}){\varphi_l}(t)}}\right]_B} ~.
\end{equation}
Equation~\eqref{eq:generalstate001} clarifies how each element of the density matrix behaves under a Gaussian noise dephasing process. In the following we will discuss two distinct scenarios regarding the environment itself: (i) the case of $N$ qubits embedded in a common bath and (ii) the case of $N$ baths completely uncorrelated.

When considering a common environment, i.e., for ${\varphi_l}(t) = \varphi(t)$ $\forall l\in\{1,2,\ldots,N\}$, the role of the coherence orders becomes explicit and the decay rates are unique for each order. Indeed, first notice that for the common environment scenario Eq.~\eqref{eq:auxfunction0001} becomes
\begin{equation}
{\xi_{{j_1}{j_2}\ldots{j_N}}} = {\left[{e^{-i({n_+} - {n_-}){\varphi}(t)}}\right]_B} ~.
\end{equation}
By using Eq.~\eqref{eq:h1q001} and the definition of ${\varphi}(t)$, we get
\begin{equation}
\label{eq:media004common002}
{\xi_{{j_1}{j_2}\ldots{j_N}}} = {e^{-im{\omega_0}t}}{e^{-{m^2}{\lambda^2}\beta(t)}} ~.
\end{equation}
Substituting Eq.~\eqref{eq:media004common002} into Eq.~\eqref{eq:generalstate001}, one may verify that the evolved state can be written as follows
\begin{align}
\label{eq:common000002}
\rho(t) &= {\sum_{{j_1},{j_2},\ldots,{j_N}}}{a_{{j_1}{j_2}\ldots{j_N}}}e^{-im{\omega_0}t}{e^{-{m^2}{\lambda^2}\beta(t)}}\bigotimes_{l = 1}^N{I_{j_l}}\nonumber \\ 
&= {\sum_{m = -N}^{N}}{e^{-im{\omega_0}t}}{e^{-{m^2}{\lambda^2}\beta(t)}}{\sum_{{n_+} - {n_-} = m}}\, {a_{{j_1}{j_2}\ldots{j_N}}} \, \bigotimes_{l = 1}^{N}{I_{j_l}} ~.
\end{align}
Given the previous results, the $\ell_1$ norm of each coherence order is given by
\begin{equation}
\label{coh_common}
{\mathcal{C}_{\vert m\vert}^{\ell_1}}(\rho(t)) = {e^{-{m^2}{\lambda^2}\beta(t)}} \, {\mathcal{C}_{m}^{\ell_1}}(\rho(0)) ~.
\end{equation}
Equation~\eqref{coh_common} implies that as long as $m$ increases, the $\ell_1$ norm of each coherence order decays faster and the effects of the bath on the system become more severe. Moreover, for $m = 0$ there is no decoherence and thus such decoherence-free subspaces can be used to encode logical qubits as a passive way to perform quantum error correction tasks~\cite{decohFS1,decohFS2}. These subspaces have been realized in several experimental platforms, e.g., NMR~\cite{NMRdecohFS1,NMRdecohFS2} and trapped ions~\cite{trapdecohFS1,trapdecohFS2,trapdecohFS3}. To illustrate these results, we assume the stochastic field $B(t)$ in Eq.~\eqref{eq:h1q0010} to be driven by the Ornstein-Uhlenbeck noise~\cite{PARIS2014256,BENEDETTI20142495,S0219749915600035}, described by the autocorrelation function ${K_{\text{OU}}}({t} - {t'},\gamma,\Gamma) = \frac{\Gamma\gamma}{2}{e^{-{\gamma}|{t} - {t'}|}}$, where $\gamma = 1/{\tau}$ plays the role of a memory noise parameter, $\tau$ is the correlation time of the process, and $\Gamma$ is the damping rate that we assume fixed. By inserting the previous autocorrelation function into Eq.~\eqref{eq:h1q0000022222}, we get ${\beta_{\text{OU}}}(t) = \frac{\Gamma}{2\gamma^2}(\gamma{t} + {e^{-\gamma{t}}} - 1)$. The behavior for a common environment is shown in Fig.~\ref{fig:commonfigure02}(a).

Let us now consider the case of $N$ independent environments, i.e., each qubit is coupled to its own environment described by uncorrelated stochastic fields ${B_l}(t)$. Therefore, Eq.~\eqref{eq:auxfunction0001} reads as
\begin{align}
\label{eq:meidnaa001}
{\xi_{{j_1}{j_2}\ldots{j_N}}} &= {\prod_{l=1}^N}\, {\left[{e^{-i({\delta_{+{j_l}}} - {\delta_{-{j_l}}})
{\varphi_l}(t)}}\right]_B} \nonumber\\
&= {\prod_{l=1}^N}\,{e^{-i({\delta_{+{j_l}}} - {\delta_{-{j_l}}}){\omega_0}t}}\, {\left[{e^{-i({\delta_{+{j_l}}} 
- {\delta_{-{j_l}}}){\lambda_l}{h_l}(t)}}\right]_B} \nonumber\\
&= {e^{-i({n_+} - {n_-}){\omega_0}t}}\, {\prod_{l=1}^N}\,{e^{-{({\delta_{+{j_l}}} - {\delta_{-{j_l}}})^2}\lambda_{l}^{2}\beta(t)}} ~.
\end{align}
Because the Kronecker $\delta$ function fulfills both properties ${\delta_{{\pm}{j_l}}^2} = {\delta_{{\pm}{j_l}}}$, and ${\delta_{+{j_l}}}{\delta_{-{j_l}}} = 0$ for all ${j_l} \in \{0,+,-,z\}$, it can be shown that ${({\delta_{+{j_l}}} - {\delta_{-{j_l}}})^2} = {\delta_{+{j_l}}} + {\delta_{-{j_l}}}$. By plugging this result into Eq.~\eqref{eq:meidnaa001}, we get
\begin{equation}
{\xi_{{j_1}{j_2}\ldots{j_N}}}  = {e^{-im{\omega_0}t}}{e^{-{\sum_{l = 1}^N}({\delta_{+{j_l}}} + {\delta_{-{j_l}}})\lambda_{l}^{2}\beta(t)}} ~.
\end{equation}
Thus, the evolved state can be written as follows
\begin{equation}
\label{eq:commondeco001}
{\rho}(t) = {\sum_{{j_1},{j_2},\ldots,{j_N}}}{a_{{j_1}{j_2}\ldots{j_N}}}\, {e^{-i m {\omega_0}t}}
{e^{-{\sum_{l = 1}^N}({\delta_{+{j_l}}} + {\delta_{-{j_l}}})\lambda_{l}^{2}\beta_l(t)}}\,{\bigotimes_{l = 1}^N}\,{I_{j_l}}    ~.
\end{equation}

Equation~\eqref{eq:commondeco001} shows that each element of the density matrix decoheres with a rate given by a linear combination of the strengths of the couplings between each qubit and its own bath, leading to a faster decay as $n_+$ and $n_-$ increase. Basically, matrix elements closer to the antidiagonal decay faster. For example, the Bell states $|{\phi^-}\rangle$ and $|{\phi^+}\rangle$, which only have zero-th and second coherence order, respectively, decohere with the same rate ${\lambda^2} = {\lambda_1^2} + {\lambda_2^2}$. Such behavior is shown in Fig.~\ref{fig:commonfigure02}(b).

\section{Hilbert-Schmidt norm and quantum metrology}

Labeling the subspaces according to coherence orders is particularly useful when dealing with phase or frequency estimation. In the last decade, it has been shown how quantum correlations can improve precision, especially through entanglement~\cite{Giovannetti1330,Maccone1,Maccone2,Maccone3,pezze,ModiPRX,Datta2012,InterPower}. The simplest metrological scenario consists of the estimation of an unknown phase acquired by a quantum system, regarded as the probe, through the interaction with another quantum system of interest. When considering the dynamics of a closed quantum system with Hamiltonian $H$, such interaction can be described by the action of the unitary operator ${U_\theta} = {e^{-i\theta H}}$ imprinting a phase shift $\theta$ on the probe state $\rho_0$, i.e., ${\rho_{\theta}} = {U_{\theta}}\, {\rho_0}{U^{\dagger}_{\theta}}$.

The quantum Fisher information (QFI), a widely applied figure of merit for quantum estimation, provides a distinguishability measure of the neighboring states ${\rho_{\theta}}$ and ${\rho_{\theta + \delta\theta}}$ when changing the phase shift $\theta$ by an infinitesimal amount $\delta\theta$~\cite{2009_Int_J_Quantum_Inform_7_125_Paris}. It defines a geometric distance between quantum states, since one can prove that QFI is related to the Bures angle, i.e., a Riemannian metric defined over the space of quantum states~\cite{Braunstein1994,Brody1996,Brody2011}. Let us consider that $\theta$ has been encoded on the initial state $\rho_0$ via an arbitrary dynamics. Given the spectral decomposition ${\rho_{\theta}} = {\sum_j}\,{q_j}|{\psi_j}\rangle\langle{\psi_j}|$ of the final state, then the QFI of ${\rho_{\theta}}$ when estimating the parameter $\theta$ can be defined as
\begin{equation}
\label{eq:QFisher}
{F_Q}({\rho_{\theta}}) = \frac{1}{2}\, {\sum_{n, m = 1}^d}\, \frac{{|\langle{\psi_n}\vert {\partial_{\theta}}\,{\varrho_{\theta}}|{\psi_m}\rangle |}^2}{{q_n} + {q_m}} ~,
\end{equation}
where ${\partial_{\theta}} \equiv \partial/\partial\theta$, $d = {2^N}$, with $N$ being the number of qubits, and the sum runs over the pair of labels $\{m,n\}$ related to the set of eigenvalues satisfying ${q_m} + {q_n} \neq 0$. The QFI (i) is additive, i.e., if the evolved state is a product one ${\rho_{\theta}} \equiv {\rho_{\theta}^{\otimes N}}$, then the QFI fulfills ${F_Q}({\rho_{\theta}^{\otimes N}}) = N {F_Q}({\rho_{\theta}})$; and (ii) reduces to the Fubini-Study metric ${F_Q}({\rho_{\theta}}) = \langle{\partial_{\theta}}{\psi_{\theta}}|{\partial_{\theta}}{\psi_{\theta}}\rangle - {|\langle{\partial_{\theta}}{\psi_{\theta}}|{\psi_{\theta}}\rangle|^2}$ if ${\rho_{\theta}} = |{\psi_{\theta}}\rangle\langle{\psi_{\theta}}|$ is pure~\cite{Quantum_1_27_2017}. In particular, if the initial state ${\rho_0}$ is a pure state undergoing an unitary evolution via ${U_\theta} = {e^{-i\theta H}}$, then the latter condition becomes simpler since QFI reduces to the variance of the generator $H$.

In this scenario, following the recent work of G\"{a}rttner, Hauke, and Rey (GHR)~\cite{A_Rey_arxiv_1706.01616}, a particularly useful coherence quantifier is the so-called MQI, defined with the Hilbert-Schmidt norm
\begin{equation}
\label{Im}
{I_m}(\rho) = \text{Tr}({\rho_{-m}}\,{\rho_m}) ~.
\end{equation}
It has been proved that MQIs define a lower bound to the quantum Fisher information, ${F_Q}({\rho_{\theta}}) \geq {F_I}(\rho, H) = {\sum_{m = 0}^N}\,F_{I}^{m}(\rho, H)$, with ${F_{I}^{m}}(\rho, H) := m^2{I_m}({\rho})$ \cite{A_Rey_arxiv_1706.01616}. It is worth noting that Eq.~\eqref{eq:QFisher} includes an additional normalization factor $1/4$ when compared to the QFI defined by GHR in Ref.~\cite{A_Rey_arxiv_1706.01616}. Each $I_m$ defines an entanglement witness for genuinely multipartite entanglement and takes into account the contributions of both $\pm m$th orders~\cite{A_Rey_arxiv_1706.01616}. Notice that one can write Eq.~\eqref{Im} as the squared Hilbert-Schmidt norm discussed in Sec.~\ref{sec:measurescoherence002}.

Let us analyze the role played by quantum coherence into the speed of a quantum state~\cite{PhysRevA.96.042327}, i.e., its rate of change when undergoing the action of a global phase shift. In summary, the main idea relies on proving that such a rate can be described as a function of the MQI defined in Eq.~\eqref{Im}~\cite{A_Rey_nature_13_781,A_Rey_arxiv_1706.01616}. The squared speed is defined as
\begin{equation}
\label{eq:despnew0000001}
{\mathcal{S}_{\tau}}({\rho_0},H) := \frac{1}{\tau^2}\left[ {\langle{\rho_0}\rangle_{\rho_0}} - {\langle{\rho_{\tau}}\rangle_{\rho_0}} \right] ~,
\end{equation}
with ${\langle\bullet\rangle_{\rho_0}} = \text{Tr}(\bullet\,{\rho_0})$ and ${\rho_{\tau}} = {U_{\tau}}\,{\rho_0}{U_{\tau}^{\dagger}}$. As pointed out by Zhang {\it et al}.~\cite{PhysRevA.96.042327}, the squared speed is positive, i.e., ${\mathcal{S}_{\tau}}({\rho_0},H) \geq 0$, and also upper bounded by the quantum Fisher information, ${\mathcal{S}_{\tau}}({\rho_0},H) \leq {F_Q}({\rho_{\tau}})$, $\forall {\rho_0},\tau,H$. Note that Eq.~\eqref{eq:despnew0000001} can be seen as a particular quantifier belonging to the family of statistical speed of quantum states recently proposed by Gessner and Smerzi~\cite{2017_arxiv_1712.04661}. Particularly, the squared speed has the advantage that the contribution of each element of the density matrix can be assessed separately, allowing us to describe how much an individual coherence order affects ${\mathcal{S}_{\tau}}({\rho_0},H)$. To see this, using Eq.~\eqref{smartdm}, we may write down the decomposition of ${\rho_0}$ into coherence orders as
\begin{equation}
\label{eq:predanadona}
{\langle{\rho_{\alpha}}\rangle_{\rho_0}} = {\sum_{m=0}^{N}}\,[{e^{-im\alpha}}{I_m}({\rho_0}) + {e^{im\alpha}}{I_{-m}}({\rho_0})] ~,
\end{equation}
with $\alpha = 0,\tau$. When plugging Eq.~\eqref{eq:predanadona} into Eq.~\eqref{eq:despnew0000001}, we conclude that the squared speed is related to the MQI as follows
\begin{equation}
\label{danadona}
{\mathcal{S}_{\tau}}({\rho_0},H) = {\sum_{m = 0}^N}\,{\mathcal{B}_{\tau,m}}({\rho_0},H) ~,
\end{equation}
where we define 
\begin{equation}
\label{danadona0004}
{\mathcal{B}_{\tau,m}}({\rho_0},H) := \frac{2}{\tau^2} \left[1 - \cos({m\tau}) \right] {I_m}({\rho_0}) ~.
\end{equation}
Therefore, one may realize that the presence of the MQI in Eq.~\eqref{danadona0004} allows us to describe how much an individual coherence order affects the speed ${\mathcal{S}_{\tau}}({\rho_0},H)$. We emphasize that each term in the summation in the right-hand side of Eq.~\eqref{danadona} is a positive real number. Furthermore, when choosing $m = {m_{\text{max}}}$ as describing the maximum nonzero coherence order of the probe state, with $0 \leq {m_{\text{max}}} \leq N$, we conclude that
\begin{equation}
\label{eq:desp0000int0005}
{\mathcal{S}_{\tau}}({\rho_0},H) \geq {\mathcal{B}_{\tau,{m_{\text{max}}}}}({\rho_0},H) ~.
\end{equation}
Since the squared speed is upper bounded by the quantum Fisher information, we have
\begin{equation}
{\mathcal{B}_{\tau,{m_{\text{max}}}}}({\rho_0},H) \leq {\mathcal{S}_{\tau}}({\rho_0},H) \leq {F_Q}({\rho_{\tau}}) ~. 
\end{equation}
This inequality is saturated for pure states in the limit $\tau\rightarrow 0$, provided the state only has coherence on the $m_{\text{max}}$-th order. Moreover, such a bound is robust to different definitions of the QFI~\cite{PhysRevA.96.042327}. 

Let us compare the different bounds on the QFI given by ${S_{\tau}}({\rho_0},H)$ and ${F_I}({\rho_0},H)$. In particular, if $x \leq \pi/2$ thus $1 - \cos{x} \geq 4{x^2}/{\pi^2}$. Therefore, when respecting the constraint $m\tau \leq \pi/2$, after some simple calculations one obtains the bound ${\mathcal{S}_{\tau}}({\rho_0},H) \geq ({8}/{\pi^2})\,{F_I}({\rho_0},H)$. Since the squared speed is upper bounded by the QFI, we get the following chain of inequalities 
\begin{equation}
\label{eq:desp0000int00011}
\frac{8}{\pi^2}\,{F_I}({\rho_0},H) \leq {\mathcal{S}_{\tau}}({\rho_0},H) \leq {F_Q}({\rho_{\tau}}) ~.
\end{equation}
This bound in Eq.~\eqref{eq:desp0000int00011} differs by a factor of $2/{\pi^2}$ from that reported by G\"{a}rttner, Hauke, and Rey~\cite{A_Rey_arxiv_1706.01616}. 
It is significant that this bound also holds when addressing each coherence order separately, i.e., for ${\mathcal{B}_{\tau,{m_{\text{max}}}}}$ and ${F_I^{m_{\text{max}}}}$.

For phase estimation, the parameter $\tau$ would be imprinted on the probe state by means the generator $H = {\sum_{l=1}^N}\,{\mathbb{I}^{\otimes{l - 1}}}\otimes{I_z}\otimes{\mathbb{I}^{\otimes{N - l}}}$. The parameter $\tau$ is estimated through an unbiased estimator $\hat\tau$ and its precision is bounded by the QFI, ac\-cor\-ding to the quantum Cram\'{e}r-Rao bound~\cite{Braunstein1994}. For separable states, it is known the QFI exhibits a linear dependence in the number of qubits of the probe system, while some entangled states show a quadratic scaling. According to Pezz\'{e} and Smerzi~\cite{pezze}, the QFI is able to detect entanglement if ${F_Q}({\rho_{\tau}}) > N/4$. We stress that such a bound differs from a factor of $4$ to that one derived in Ref.~\cite{pezze} due to the normalization adopted to the QFI in Eq.~\eqref{eq:QFisher}. Since both ${\mathcal{B}_{\tau,{m_{\text{max}}}}}({\rho_0},H)$ and ${F_I^m}({\rho_0}, H)$ characterize lower bounds to the QFI, it is straightforward to obtain a simple criterion to testify quantum enhanced precision by following Ref.~\cite{pezze}. We gua\-ran\-tee a quantum advantage if
\begin{equation}
\label{tapiocas}
{\mathcal{B}_{\tau,{m_{\text{max}}}}}({\rho_0},H) > \frac{N}{4} \quad \text{or} \quad {F_I^{m_{\text{max}}}}({\rho_0}, H) > \frac{N}{4} ~.
\end{equation}

In comparison to any figure of merit in the literature, both inequalities in Eq.~\eqref{tapiocas} require much less information from $\rho_0$ to evaluate their usefulness for phase estimation. While the squared speed or even ${F_I}({\rho_0},{H})$ rely on information from the whole density matrix, these bounds only depend on a minimal set of elements whose cardinality given by Eq.~\eqref{mcohnumber}. Therefore, in order to get a quantum advantage in phase estimation, one just needs to maximize the highest coherence order attainable within the control limitations of a given experimental setup. For example, the highest order in a three-qubit system can be the second order, due to the lack of universal control over the qubits or decoherence effects during state preparation. This has been addressed by the NMR community under different experimental conditions~\cite{Levitt1,Levitt2,GlaserJCP2016}. Moreover, for a set of states $\{\rho^{(m)}\}$, $m \in\{0, 1, \ldots, N\}$, all with the same amount of coherence just on the $m$th coherence order, the Fisher information follows the rule ${F_Q}(\rho^{(0)})\leq {F_Q}(\rho^{(1)})\leq\ldots\leq {F_Q}(\rho^{(N-1)})\leq {F_Q}(\rho^{(N)})$. The quantum advantage for phase estimation is only achieved for states with $m_{\text{max}} \geq 2$.

Entanglement is considered to be the key ingredient to achieve precision beyond the best classical strategy. In particular, the GHZ state, $|{\text{GHZ}_N}\rangle = \frac{1}{\sqrt{2}}\left(\,{|0\rangle^{\otimes N}} + {|1\rangle^{\otimes N}}\right)$, is pointed to as achieving the so-called Heisenberg limit, when ${F_Q} \propto {N^2}$. However, only entanglement is not enough~\cite{PRAallentang}, and Eq.~\eqref{danadona} makes it clear. Let us focus on the three-qubit case, i.e., $N = 3$. On the one hand, the maximally entangled state, $|\text{W}\rangle = \frac{1}{\sqrt{3}}(|{001}\rangle + |{010}\rangle + |100\rangle)$, cannot provide any information on a phase shift generated by $Z$ since it is invariant under such transformation, i.e., ${S_{\tau}}({|\text{W}\rangle\langle\text{W}|}) = {\mathcal{B}_{\tau,{m_{\text{max}}}}}({|\text{W}\rangle\langle\text{W}|}) = 0$. On the other hand, the GHZ state has only the highest coherence order, i.e., $N$. Basically, the way in which information is encoded over the coherence orders tells us how useful a state is for phase estimation. As another example, both states $|{++}\rangle = \frac{1}{2}(|00\rangle + |01\rangle + |10\rangle + |11\rangle)$ and $|\varphi\rangle = \frac{1}{2}(|00\rangle + |01\rangle + |10\rangle - |11\rangle)$ give the same precision, despite the former being separable and the latter maximally entangled.
\begin{figure}[t]
\begin{center}
\begin{tikzpicture}
    \begin{groupplot}[
        group style={
            group name=my plots,
            group size=2 by 2,
            xlabels at=edge bottom,
            ylabels at=edge left,
            vertical sep=1.5cm,
        },
         trig format plots = rad,
        footnotesize,
        width=5cm,
        height=4.5cm,
        xlabel= $p$,
        enlargelimits=false,
        xmin=0, xmax=1,
         tick style = {thick},
         scaled ticks=true,
     tick scale binop=\times,
      legend columns=3,
    legend entries={[black]${F_Q}({\rho_{\tau}})$~~~~~\\[black]${\mathcal{S}_{\tau}}({\rho_0},H)$~~~~~\\[black]${\mathcal{B}_{{\tau},{m_{\text{max}}}}}({\rho_0},H)$~~~~~\\[black]${F_Q}({\rho_{\tau}^{\text{class}}})$~~~~~\\[black]${F_I}({\rho_0},H)$~~~~~\\[black]${F_I^{\,{m_{\text{max}}}}}({\rho_0},H)$\\},
   legend to name=namedgroups01XX,
   legend style={draw=none,font=\footnotesize},
   legend cell align=center
]
    \nextgroupplot[title={{{$\phi = 0$}}}, xlabel= $p$]
\addplot[
   red,
   very thick,
   domain=0:1,
   samples=300,
]
   {((9*x*x)/(1+3*x))};
   
   
   \addplot[
   blue,
   dashed,
   very thick,
   domain=0:1,
   samples=300,
]
   {(1/(2*(pi/6)*(pi/6)))*x*x*cos(0)*cos(0)*sin(pi/12)*sin(pi/12)*(9 - 3*cos(2*0) + 8*cos(pi/6) + 4*cos(0)*cos(0)*cos(2*pi/6))};
   
        \addplot[
   black,
   very thick,
    dashdotted,
   domain=0:1,
   samples=300,
]
   {(1/((pi/6)*(pi/6)))*x*x*cos(0)*cos(0)*cos(0)*cos(0)*sin(pi/4)*sin(pi/4)};
   
   
   \addplot[
   gray,
   thick,
   dotted,
   domain=0:1,
   samples=300,
]
   {0.75};
   
   
           \addplot[
   brown,
   very thick,
   loosely dotted,
   domain=0:1,
   samples=300,
]
   {(9/4)*x*x};
   
   
              \addplot[
   magenta,
   very thick,
   loosely dashdotted,
   domain=0:1,
   samples=300,
]
   {(9/4)*x*x};
   
   
   \node at (axis cs:0.2,1.99) {{\footnotesize{(a)}}};
    \nextgroupplot[title={{{$\phi = \pi/6$}}}, xlabel= $p$, ytick = {0,0.4,0.8,1.2,1.6}]
    \addplot[
   red,
   very thick,
   domain=0:1,
   samples=300,
]
   {((111*x*x)/(16*(1+3*x)))};   
   
     
   \addplot[
   blue,
   very thick,
   dashed,
   domain=0:1,
   samples=300,
]
   {(1/(2*(pi/6)*(pi/6)))*x*x*cos(pi/6)*cos(pi/6)*sin(pi/12)*sin(pi/12)*(9 - 3*cos(2*pi/6) + 8*cos(pi/6) + 4*cos(pi/6)*cos(pi/6)*cos(2*pi/6) )};
   
   
           \addplot[
   black,
   very thick,
   dashdotted,
   domain=0:1,
   samples=300,
]
   {(1/((pi/6)*(pi/6)))*x*x*cos(pi/6)*cos(pi/6)*cos(pi/6)*cos(pi/6)*sin(pi/4)*sin(pi/4)};
   
   
      \addplot[
   gray,
   thick,
   dotted,
   domain=0:1,
   samples=300,
]
   {0.75};
   
   
           \addplot[
   brown,
   very thick,
   loosely dotted,
   domain=0:1,
   samples=300,
]
   {(111/64)*x*x};
   
   
              \addplot[
   magenta,
   very thick,
   loosely dashdotted,
   domain=0:1,
   samples=300,
]
   {(81/64)*x*x};
   
   
  \node at (axis cs:0.2,1.5) {{\footnotesize{(b)}}};
    \nextgroupplot[title={{{$\phi = \pi/4$}}}]
    \addplot[
   red,
   very thick,
   domain=0:1,
   samples=300,
]
   {((19*x*x)/(4*(1+3*x)))};
   

    \addplot[
   blue,
   dashed,
   very thick,
   domain=0:1,
   samples=300,
]
   {(1/(2*(pi/6)*(pi/6)))*x*x*cos(pi/4)*cos(pi/4)*sin(pi/12)*sin(pi/12)*(9 - 3*cos(2*pi/4) + 8*cos(pi/6) + 4*cos(pi/4)*cos(pi/4)*cos(2*pi/6))};
   
   
           \addplot[
             black,
   very thick,
    dashdotted,
   domain=0:1,
   samples=300,
]
   {(1/((pi/6)*(pi/6)))*x*x*cos(pi/4)*cos(pi/4)*cos(pi/4)*cos(pi/4)*sin(pi/4)*sin(pi/4)};
   
   
      \addplot[
   gray,
   thick,
   dotted,
   domain=0:1,
   samples=300,
]
   {0.75};

   
           \addplot[
   brown,
   very thick,
   loosely dotted,
   domain=0:1,
   samples=300,
]
   {(19/16)*x*x};
   
   
              \addplot[
   magenta,
   very thick,
   loosely dashdotted,
   domain=0:1,
   samples=300,
]
   {(9/16)*x*x};
   
   
   \node at (axis cs:0.2,1.025) {{\footnotesize{(c)}}};
    \nextgroupplot[title={{{$\phi = \pi/3$}}}, ymin=0, ymax=1]
    \addplot[
   red,
   very thick,
   domain=0:1,
   samples=300,
]
   {((39*x*x)/(16*(1+3*x)))};
   
   
    \addplot[
   blue,
   very thick,
   dashed,
   domain=0:1,
   samples=300,
]
   {(1/(2*(pi/6)*(pi/6)))*x*x*cos(pi/3)*cos(pi/3)*sin(pi/12)*sin(pi/12)*(9 - 3*cos(2*pi/3) + 8*cos(pi/6) + 4*cos(pi/3)*cos(pi/3)*cos(2*pi/6))};
   
   
           \addplot[
   black,
   very thick,
    dashdotted,
   domain=0:1,
   samples=300,
]
   {(1/((pi/6)*(pi/6)))*x*x*cos(pi/3)*cos(pi/3)*cos(pi/3)*cos(pi/3)*sin(pi/4)*sin(pi/4)};
   
   
        \addplot[
   gray,
   thick,
   dotted,
   domain=0:1,
   samples=300,
]
   {0.75};
   
   
           \addplot[
   brown,
   very thick,
   loosely dotted,
   domain=0:1,
   samples=300,
]
   {(39/64)*x*x};
   
   
              \addplot[
   magenta,
   very thick,
   loosely dashdotted,
   domain=0:1,
   samples=300,
]
   {(9/64)*x*x};
   
  \node at (axis cs:0.2,0.875) {{\footnotesize{(d)}}};
    \end{groupplot}
\end{tikzpicture}
   \\
   \ref{namedgroups01XX}
\caption{(Color online) Plot of quantum Fisher information ${F_Q}({\rho_{\tau}})$ (red solid line), squared speed ${\mathcal{S}_{\tau}}({\rho_0},H)$ (blue dashed line), ${\mathcal{B}_{{\tau},{m_{\text{max}}}}}({\rho_0},H)$ (black dot dashed line), ${F_I}({\rho_0},H)$ (brown loosely dotted line), and ${F_I^{m_{\text{max}}}}({\rho_0},H)$ (magenta loosely dot dashed line) related to the evolved state ${\rho_{\tau}} = {U_{\tau}}\,{\rho_0}{U_{\tau}^{\dagger}}$, where ${\rho_0} = \left(\frac{1 - p}{8}\right)\mathbb{I} + p\,|\Psi\rangle\langle{\Psi}|$ and $|\Psi\rangle = \cos{\phi}\,|\text{GHZ}_3\rangle + \sin{\phi}\,|{001}\rangle$ (${m_{\text{max}}}$), for (a)~$\phi = 0$, (b)~$\phi = \pi/6$, (c)~$\phi = \pi/4$, and (d)~$\phi = \pi/3$. Notice that the unitary evolution is generated by the Hamiltonian $H = {\sum_{l=1}^3}\,{\mathbb{I}^{\otimes\, {l - 1}}}\otimes{I_l^z}\otimes{\mathbb{I}^{\otimes\, {3 - l}}}$. Here we choose $\tau = \pi/6$. The gray dotted line represents the QFI related to the three-qubits uniform superposition initial state ${\rho_0^{\text{class}}} = |{{+}{+}{+}}\rangle\langle{{+}{+}{+}}|$ which undergoes such unitary dynamics.}
\label{fig:examples02}
\end{center}
\end{figure}
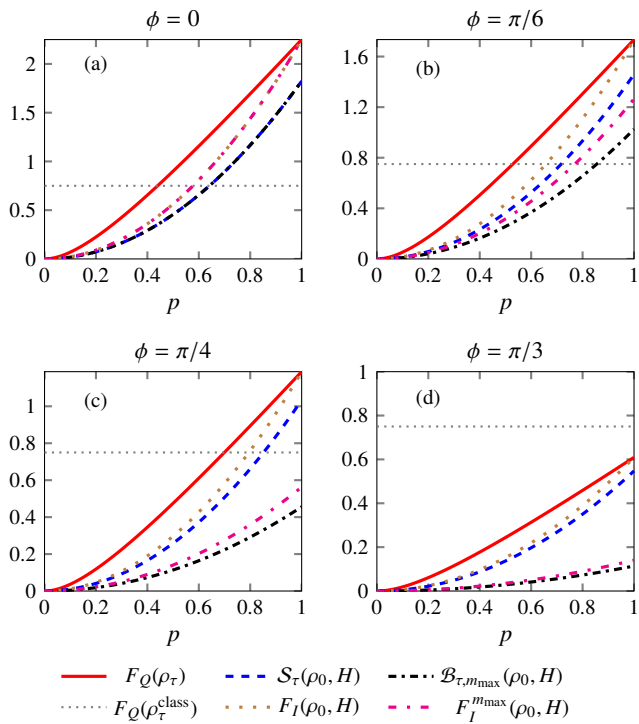

As an example, let us consider the family of states given by ${\rho_0} = \left(\frac{1 - p}{2^N}\right){\mathbb{I}_N} + p\,|\Psi\rangle\langle{\Psi}|$, with $\vert\Psi\rangle = \cos{\phi}\,|{\text{GHZ}_N}\rangle + \sin{\phi}\,{|0\rangle^{\otimes N - 1}}|1\rangle$, $0 \leq \phi \leq \pi/2$, and $0\leq p\leq 1$. Such states exhibit nonzero coherences in the first and $N$th orders, with the parameter $\phi$ weighting the contribution of each order. The degree of purity is given by $\text{Tr}\,({\rho_0^2}) = [1 + (d - 1){p^2}]/d$, with $d = {2^N}$. Particularly, when choosing $N = 3$, Fig.~\ref{fig:examples02} unveils how ${\mathcal{B}_{\tau,{m_{\text{max}}}}}({\rho_0},H)$ and ${F_I^{m_{\text{max}}}}({\rho_0}, H)$ can detect metrologically useful entanglement for different values of $\phi$, as a function of $p$. From Fig.~\ref{fig:examples02}(a)--\ref{fig:examples02}(d), it can be seen that as the contribution of the first coherence order increases, then the difference between ${\mathcal{S}_{\tau}}({\rho_0},H)$ and ${\mathcal{B}_{\tau,{m_{\text{max}}}}}({\rho_0},H)$ follows this trend and the ability to detect entanglement by ${\mathcal{B}_{\tau,{m_{\text{max}}}}}({\rho_0},H)$ is reduced. Notice that the same reasoning holds for ${F_I}({\rho_0},H)$ and ${F_I^{m_{\text{max}}}}({\rho_0},H)$. In summary, it means that Eq.~\eqref{tapiocas} defines a criterion that is only sufficient to certify the usefulness of a state $\rho$ for phase estimation, as shown in Fig.~\ref{fig:examples02}(c). We emphasize that ${F_I}({\rho_0},H)$ and ${F_I}^{m_{\text{max}}}({\rho_0},H)$ offer tighter bounds on the QFI than ${\mathcal{S}_{\tau}}({\rho_0},H)$ and ${\mathcal{B}_{\tau,{m_{\text{max}}}}}({\rho_0},H)$. While ${F_I}({\rho_0},H)$ and ${\mathcal{S}_{\tau}}({\rho_0},H)$ require the knowledge of the whole density matrix $\rho_0$, both functionals ${\mathcal{B}_{\tau,{m_{\text{max}}}}}({\rho_0},H)$ and ${F_I^{m_{\text{max}}}}({\rho_0},H)$ are related solely to the maximal coherence order and thus rely on a small set of elements of the density matrix. Such a property reduces the experimental cost and could be seen as a potential advantage to the design of new quantum technologies. The calculations and some additional examples are discussed in the Appendix.


\section{Conclusions}

In this paper we adopted an approach of defining quantifiers for each coherence order based on its specific application and meaning rather than looking for a general theory. Particularly, both quantifiers in Eqs.~\eqref{quantdef} and~\eqref{Im} can be easily applied to study decoherence or quantum phase estimation, making clear how different subspaces of the Hilbert-Schmidt space can share some properties while exhibiting a qualitatively distinct behavior from other subspaces. For open quantum systems, this allowed us to readily assess which subspaces are more or less affected by dephasing from a common environment, with the zeroth coherence order related to decoherence-free subspaces. 

Here we propose a simple and sufficient way to assess the usefulness of a given quantum state for phase estimation beyond the classical limit. The biggest advantage of using such a framework lies in the minimal experimental cost required to measure a single coherence order~\cite{Teles}. From an experimental viewpoint, Eq.~\eqref{tapiocas} implies that the problem to guarantee quantum enhanced precision consists of the maximization of the amount of coherence in the highest order attainable, given limitations on the control fields and decoherence effects. This subject has been addressed in the context of NMR experiments~\cite{Levitt1,Levitt2,GlaserJCP2016}. Finally, according to Eqs.~\eqref{coh_common} and~\eqref{tapiocas}, our results show that as $m$ increases, more useful to metrology and more affected by decoherence is a particular coherence order. In a context of noisy quantum metrology, our results suggest the existence of an optimal coherence order for frequency estimation under dephasing.


\begin{acknowledgments}
We thank Amelia, Jaqueline and Magda for the delicious daily coffee. 
The authors would like to acknowledge the financial support from the Brazilian ministries MEC and MCTIC, funding agencies CNPq, CAPES, FAPESP, and the Brazilian National Institute of Science and Technology of Quantum Information (INCT-IQ).
\end{acknowledgments}

\pagebreak
\setcounter{equation}{0}
\makeatletter
\renewcommand{\theequation}{A\arabic{equation}}

\section*{Appendix - Additional examples}


\subsection{Example 1}

Let us consider the state ${\rho_0} = |\psi\rangle\langle\psi|$, where $\vert\psi\rangle = \cos{\phi}\,|001\rangle + \sin{\phi}\,|{\text{GHZ}_3}\rangle$ with 
\begin{equation}
\label{eq:ghzstate}
|{\text{GHZ}_3}\rangle = \frac{1}{\sqrt{2}}(|{000}\rangle + |{111}\rangle) ~,
\end{equation} 
and also the three-qubit uniform superposition state ${\rho_0^{\text{class}}} = |{{+}{+}{+}}\rangle\langle{{+}{+}{+}}|$, with
\begin{equation}
\label{eq:desp00003}
|+\rangle = \frac{1}{\sqrt{2}}(|{0}\rangle + |{1}\rangle) ~.
\end{equation}
Both states ${\rho_0}$ and ${\rho_0^{\text{class}}}$ undergo the unitary evolution ${\rho_{\tau}} = {U_{\tau}}\,{\rho_0}{U_{\tau}^{\dagger}}$ and ${\rho^{\text{class}}_{\tau}} = {U_{\tau}}\,{\rho_0^{\text{class}}}\,{U_{\tau}^{\dagger}}$, respectively, where ${U_{\tau}} = {e^{-i\tau H}}$, and also
\begin{equation}
\label{eq:generatorNequal3}
H = {\sum_{l=1}^3}\,{\mathbb{I}^{\otimes\, {l - 1}}}\otimes{I_l^z}\otimes{\mathbb{I}^{\otimes\, {3 - l}}} ~,
\end{equation}
with ${I_l^z} = (1/2)\,{\sigma_l^z}$. Because the unitary evolution does not change the purity of the probe states ${\rho_0}$ and ${\rho_0^{\text{class}}}$, then both final states ${\rho_{\tau}}$ and ${\rho^{\text{class}}_{\tau}}$ will be also pure. In this case, it can be shown that the quantum Fisher information related to ${\rho_{\tau}}$ and ${\rho^{\text{class}}_{\tau}}$ reduces to the variance of the generator $H$ with respect to both probe states ${\rho_0}$ and ${\rho_0^{\text{class}}}$, respectively, i.e.,
\begin{align}
\label{eq:fisherexample1}
{F_Q}({\rho_{\tau}}) &= \langle{\psi}|{H^2}|{\psi}\rangle - {{\langle{\psi}|{H}|{\psi}\rangle}^2} \nonumber\\
&= \frac{1}{8}\,[19 + \cos(2\phi)]\,{\sin^2}\phi ~,
\end{align}
and 
\begin{align}
\label{eq:fisherexample10000b}
{F_Q}({\rho^{\text{class}}_{\tau}}) &= \langle{{+}{+}{+}}|{H^2}|{{+}{+}{+}}\rangle - {{\langle{{+}{+}{+}}|{H}|{{+}{+}{+}}\rangle}^2} \nonumber\\
&= \frac{3}{4} ~.
\end{align}

Notice that both states ${\rho_{\tau}}$ and ${\rho^{\text{class}}_{\tau}}$ have ${m_{\text{max}}} = 3$ nonzero coherence order. In this case, the coherence quantifiers ${\mathcal{S}_{\tau}}({\rho_0},H)$ and ${\mathcal{B}_{{\tau},3}}({\rho_0},H)$, defined in Eq.~\eqref{danadona}, read as 
\begin{align}
\label{eq:fisherexample102}
{\mathcal{S}_{\tau}}({\rho_0},H) = \frac{1}{2{\tau^2}} &\left[9 + 3\cos(2\phi) + 8\cos\tau \right. \nonumber\\ 
&\left. + 4\,{\sin^2}\phi\cos(2\tau)\right]\,{\sin^2}\phi\,{{\sin^2}\left(\frac{\tau}{2}\right)}
\end{align}
and
\begin{equation}
\label{eq:fisherexample103}
{\mathcal{B}_{{\tau},3}}({\rho_0},H) = \frac{1}{\tau^2}\,{\sin^4}\phi\,{\sin^2}\left(\frac{3\tau}{2}\right) ~.
\end{equation}
Interestingly, it is straightforward to conclude the following
\begin{equation}
\label{eq:fisherexample103bbbb}
{F_I}({\rho_0},H) = \frac{1}{8}\,[19 + \cos(2\phi)]\,{\sin^2}\phi 
\end{equation}
and
\begin{equation}
\label{eq:fisherexample103bbbb2}
{F_I^{m_{\text{max}}}}({\rho_0},H) = \frac{9}{4}\,{\sin^4}\phi ~.
\end{equation}
In Fig.~\ref{example1dot02}(a) we plot the QFI, $\mathcal{S_{\tau}}({\rho_0},H)$, $\mathcal{B_{\tau}}({\rho_0},H)$, ${F_I}({\rho_0},H)$, and ${F_I^{m_{\text{max}}}}({\rho_0},H)$ given in Eqs.~\eqref{eq:fisherexample1},~\eqref{eq:fisherexample10000b},~\eqref{eq:fisherexample102},~\eqref{eq:fisherexample103},~\eqref{eq:fisherexample103bbbb}, and~\eqref{eq:fisherexample103bbbb2}, respectively.

\subsection{Example 2}

Let us consider the initial pure state ${\rho_0} = |\psi\rangle\langle\psi|$, with $\vert\psi\rangle = \cos{\phi}\,|000\rangle + \sin{\phi}\,|\text{W}\rangle$ and 
\begin{equation}
|\text{W}\rangle = \frac{1}{\sqrt{3}}(|011\rangle + |101\rangle + |110\rangle) ~.
\end{equation}
Analogous to the previous example, such a probe state undergoes the unitary evolution ${\rho_{\tau}} = {U_{\tau}}\,{\rho_0}{U_{\tau}^{\dagger}}$, with ${U_{\tau}} = {e^{-i\tau H}}$ and the Hamiltonian is defined in Eq.~\eqref{eq:generatorNequal3}. Because the final state ${\rho_{\tau}}$ is a pure one, then the quantum Fisher information related to ${\rho_{\tau}}$ is given by the variance of the generator $H$, i.e.,
\begin{align}
\label{eq:fisherexample2}
{F_Q}({\rho_{\tau}}) &= \langle{\psi}|{H^2}|{\psi}\rangle - {{\langle{\psi}|{H}|{\psi}\rangle}^2} \nonumber\\
&= {\sin^2}(2\phi) ~.
\end{align}
Notice that the states $\rho_0$, ${\rho_{\tau}}$ and ${\rho_0^{\text{class}}} = |{{+}{+}{+}}\rangle\langle{{+}{+}{+}}|$, with $|+\rangle$ defined in Eq.~\eqref{eq:desp00003}, have only ${m_{\text{max}}} = 2$ nonzero coherence order. Therefore, one can verify that the coherence quantifiers ${\mathcal{S}_{\tau}}({\rho_0},H)$ and ${\mathcal{B}_{{\tau},2}}({\rho_0},H)$ defined in Eq.~\eqref{danadona} become
\begin{equation}
\label{eq:fisherexample201}
\mathcal{S_{\tau}}({\rho_0},H) = {\mathcal{B}_{{\tau},2}}({\rho_0},H) =  \frac{1}{\tau^2}\,{\sin^2}(2\phi)\,{\sin^2}\tau ~.
\end{equation}
Furthermore, one can readily verify that 
\begin{equation}
\label{eq:fisherexample201bbbbbb}
{F_I}({\rho_0},H) = {F_I^{m_{\text{max}}}}({\rho_0},H) = {\sin^2}(2\phi) ~.
\end{equation}
In Fig.~\ref{example1dot02}(b) we plot the QFI for both states ${\rho_{\tau}}$ and ${\rho_{\tau}^{\text{class}}}$, $\mathcal{S_{\tau}}({\rho_0},H)$, $\mathcal{B_{\tau}}({\rho_0},H)$, ${F_I}({\rho_0},H)$, and ${F_I^{m_{\text{max}}}}({\rho_0},H)$ given in Eqs.~\eqref{eq:fisherexample2},~\eqref{eq:fisherexample10000b},~\eqref{eq:fisherexample201}, and~\eqref{eq:fisherexample201bbbbbb}, respectively.
\begin{figure}[t]
\begin{tabular}{rr}
\begin{tikzpicture}[scale=0.9,font=\normalsize]
\begin{axis}[
   trig format plots = rad,
   enlargelimits=false,
   very thick,
   xlabel={$\phi$},
   xtick = {0,0.785398,1.5708,2.35619,3.14159},
   xticklabels = {$0$,$\frac{\pi}{4}$,$\frac{\pi}{2}$,$\frac{3\pi}{4}$,$\pi$},
   ymin=0, ymax=2.3,
    legend columns=3,
    legend entries={[black]${F_Q}({\rho_{\tau}})$~~~~~\\[black]${\mathcal{S}_{\tau}}({\rho_0},H)$~~~~~\\[black]${\mathcal{B}_{{\tau},{m_{\text{max}}}}}({\rho_0},H)$~~~~~\\[black]${F_Q}({\rho_{\tau}^{\text{class}}})$~~~~~\\[black]${F_I}({\rho_0},H)$~~~~~\\[black]${F_I^{\,{m_{\text{max}}}}}({\rho_0},H)$\\},
    legend to name=namedexample1,
    legend style={draw=none,font=\small},   
   legend cell align=center,
   tick style = {very thick}
]
\addplot[
   red,
   very thick,
   domain=0:pi,
   samples=300,
]
   {(1/8)*(19 + cos(2*x))*sin(x)*sin(x)};

\addplot[
   blue,
   very thick,
   dashed,
   domain=0:pi,
   samples=300,
]
   {( 1/(2*(pi/6)*(pi/6)) )*sin(x)*sin(x)*sin(pi/12)*sin(pi/12)*( 9 + 3*cos(2*x) + 8*cos(pi/6) + 4*cos(2*pi/6)*sin(x)*sin(x) )};

\addplot[
   black,
   very thick,
   dashdotted,
   domain=0:pi,
   samples=300,
]
   {(1/((pi/6)*(pi/6)))*sin(x)*sin(x)*sin(x)*sin(x)*sin(pi/4)*sin(pi/4)};

\addplot[
   gray,
   very thick,
   dotted,
   domain=0:pi,
   samples=300,
]
   {0.75};
   
   
           \addplot[
   brown,
   very thick,
   loosely dotted,
   domain=0:pi,
   samples=300,
]
   {(1/8)*(19 + cos(2*x))*sin(x)*sin(x)};
   
   
              \addplot[
   magenta,
   very thick,
   loosely dashdotted,
   domain=0:pi,
   samples=300,
]
   {(9/4)*sin(x)*sin(x)*sin(x)*sin(x)};
  
\node at (axis cs:0.392699,2.1) {{\large{(a)}}};
\end{axis}
\end{tikzpicture}
\\
\quad\quad
\begin{tikzpicture}[scale=0.9,font=\normalsize]
\begin{axis}[
   trig format plots = rad,
   enlargelimits=false,
   very thick,
   xlabel={$\phi$},
   xtick = {0,0.785398,1.5708,2.35619,3.14159},
   xticklabels = {$0$,$\frac{\pi}{4}$,$\frac{\pi}{2}$,$\frac{3\pi}{4}$,$\pi$},
   ymin=0, ymax=1.1,
   tick style = {very thick}
]
\addplot[
   red,
   very thick,
   domain=0:pi,
   samples=300,
]
   {sin(2*x)*sin(2*x)};

\addplot[
   blue,
   very thick,
   dashed,
   domain=0:pi,
   samples=300,
]
   {(1/((pi/6)*(pi/6)))*sin(2*x)*sin(2*x)*sin(pi/6)*sin(pi/6)};
   
     \addplot[
   black,
   very thick,
    dashdotted,
   domain=0:pi,
   samples=300,
]
   {(1/((pi/6)*(pi/6)))*sin(2*x)*sin(2*x)*sin(pi/6)*sin(pi/6)};

\addplot[
   gray,
   thick,
   dotted,
   domain=0:pi,
   samples=300,
]
   {0.75};
   
   
           \addplot[
   brown,
   very thick,
   loosely dotted,
   domain=0:pi,
   samples=300,
]
   {sin(2*x)*sin(2*x)};
   
   
              \addplot[
   magenta,
   very thick,
   loosely dashdotted,
   domain=0:pi,
   samples=300,
]
   {sin(2*x)*sin(2*x)};
\node at (axis cs:0.392699,1) {{\large{(b)}}};
\end{axis}
\end{tikzpicture}
\end{tabular}
\\
   \ref{namedexample1}
\caption{(Color online) Plot of quantum Fisher information ${F_Q}({\rho_{\tau}})$ (red solid line), squared speed ${\mathcal{S}_{\tau}}({\rho_0},H)$ (blue dashed line), ${\mathcal{B}_{{\tau},{m_{\text{max}}}}}({\rho_0},H)$ (black dot dashed line), ${F_I}({\rho_0},H)$ (brown loosely dotted line), and ${F_I^{m_{\text{max}}}}({\rho_0},H)$ (magenta loosely dot dashed line) related to the evolved state ${\rho_{\tau}} = {U_{\tau}}\,{\rho_0}{U_{\tau}^{\dagger}}$, where (a)~example 1: ${\rho_0} = |\psi\rangle\langle\psi|$, with $\vert\psi\rangle = \cos{\phi}\,|001\rangle + \sin{\phi}\,|{\text{GHZ}_3}\rangle$ (${m_{\text{max}}} = 3$), and (b)~example 2: ${\rho_0} = |\psi\rangle\langle\psi|$, with $\vert\psi\rangle = \cos{\phi}\,|000\rangle + \sin{\phi}\,|\text{W}\rangle$ (${m_{\text{max}}} = 2$). In both cases such unitary evolution is generated by the Hamiltonian $H$ in Eq.~\eqref{eq:generatorNequal3} and $\tau = \pi/6$. The gray dotted line represents the QFI related to the three-qubits uniform superposition initial state ${\rho_0^{\text{class}}} = |{{+}{+}{+}}\rangle\langle{{+}{+}{+}}|$ which undergoes the unitary dynamics discussed before.}
\label{example1dot02}
\end{figure}

\subsection{Example 3}

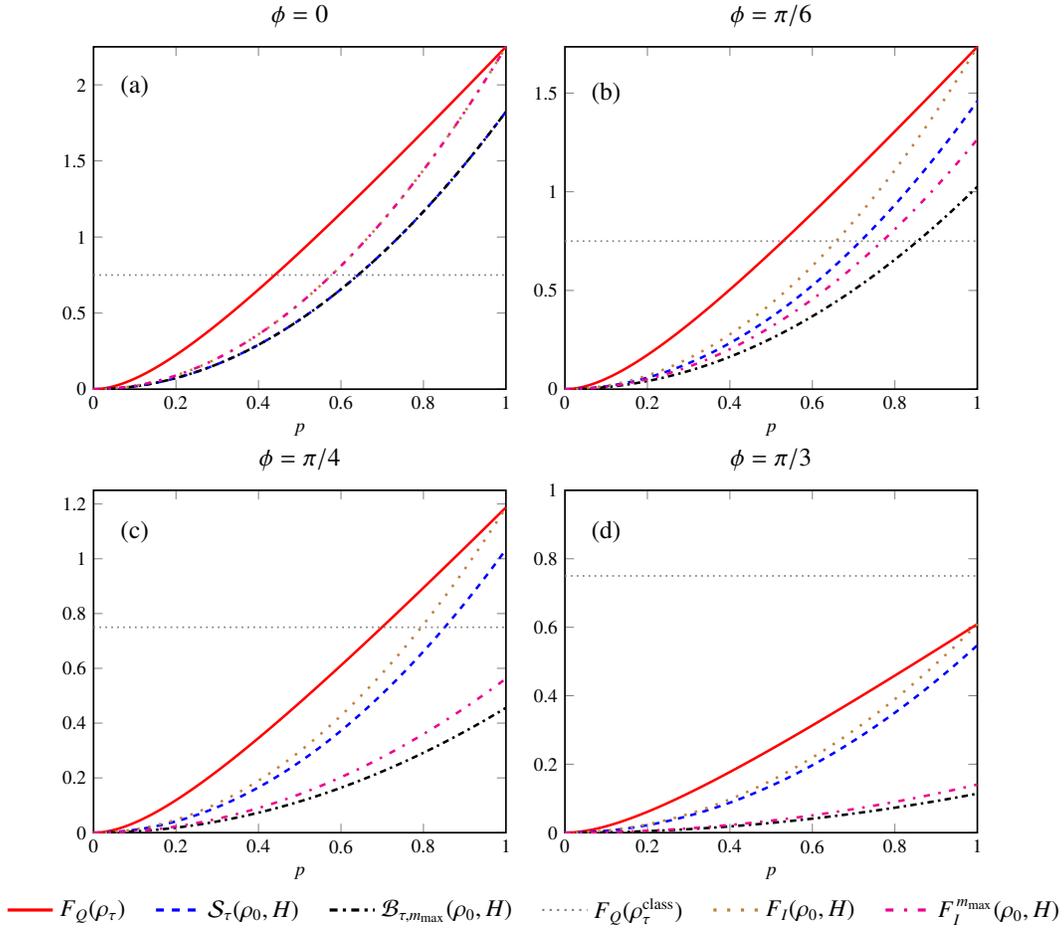
\begin{figure*}[t]
\begin{center}
\begin{tabular}{lr}
\begin{tikzpicture}[baseline,trim axis left,scale=0.8]
\begin{axis}[
   trig format plots = rad,
   title={{\large{$\phi = 0$}}},
   enlargelimits=false,
   thick,
   xmin=0, xmax=1,
    ymin=0, ymax=2.25,
   xlabel={$p$},
   legend columns=6,
    legend entries={[black]${F_Q}({\rho_{\tau}})$~~~~~\\[black]${\mathcal{S}_{\tau}}({\rho_0},H)$~~~~~\\[black]${\mathcal{B}_{{\tau},{m_{\text{max}}}}}({\rho_0},H)$~~~~~\\[black]${F_Q}({\rho_{\tau}^{\text{class}}})$~~~~~\\[black]${F_I}({\rho_0},H)$~~~~~\\[black]${F_I^{\,{m_{\text{max}}}}}({\rho_0},H)$\\},
   legend to name=named222,
   legend style={draw=none,font=\small},
   legend cell align=center
]
\addplot[
   red,
   very thick,
   domain=0:1,
   samples=300,
]
   {((9*x*x)/(1+3*x))};
   
   
   \addplot[
   blue,
   dashed,
   very thick,
   domain=0:1,
   samples=300,
]
   {(1/(2*(pi/6)*(pi/6)))*x*x*cos(0)*cos(0)*sin(pi/12)*sin(pi/12)*(9 - 3*cos(2*0) + 8*cos(pi/6) + 4*cos(0)*cos(0)*cos(2*pi/6))};
   
        \addplot[
   black,
   very thick,
    dashdotted,
   domain=0:1,
   samples=300,
]
   {(1/((pi/6)*(pi/6)))*x*x*cos(0)*cos(0)*cos(0)*cos(0)*sin(pi/4)*sin(pi/4)};
   
   
   \addplot[
   gray,
   thick,
   dotted,
   domain=0:1,
   samples=300,
]
   {0.75};
   
   
           \addplot[
   brown,
   very thick,
   loosely dotted,
   domain=0:1,
   samples=300,
]
   {(9/4)*x*x};
   
   
              \addplot[
   magenta,
   very thick,
   loosely dashdotted,
   domain=0:1,
   samples=300,
]
   {(9/4)*x*x};
   
\node at (axis cs:0.1,1.99) {{\large{(a)}}};
\end{axis}
\end{tikzpicture}
&
\begin{tikzpicture}[baseline,trim axis right,scale=0.8]
\begin{axis}[
   trig format plots = rad,
   title={{\large{$\phi = \pi/6$}}},
   enlargelimits=false,
   thick,
   xmin=0, xmax=1,
    ymin=0, ymax=1.73438,
   xlabel={$p$}
]
\addplot[
   red,
   very thick,
   domain=0:1,
   samples=300,
]
   {((111*x*x)/(16*(1+3*x)))};   
   
     
   \addplot[
   blue,
   very thick,
   dashed,
   domain=0:1,
   samples=300,
]
   {(1/(2*(pi/6)*(pi/6)))*x*x*cos(pi/6)*cos(pi/6)*sin(pi/12)*sin(pi/12)*(9 - 3*cos(2*pi/6) + 8*cos(pi/6) + 4*cos(pi/6)*cos(pi/6)*cos(2*pi/6) )};
   
   
           \addplot[
   black,
   very thick,
   dashdotted,
   domain=0:1,
   samples=300,
]
   {(1/((pi/6)*(pi/6)))*x*x*cos(pi/6)*cos(pi/6)*cos(pi/6)*cos(pi/6)*sin(pi/4)*sin(pi/4)};
   
   
      \addplot[
   gray,
   thick,
   dotted,
   domain=0:1,
   samples=300,
]
   {0.75};
   
   
           \addplot[
   brown,
   very thick,
   loosely dotted,
   domain=0:1,
   samples=300,
]
   {(111/64)*x*x};
   
   
              \addplot[
   magenta,
   very thick,
   loosely dashdotted,
   domain=0:1,
   samples=300,
]
   {(81/64)*x*x};
   
\node at (axis cs:0.1,1.5) {{\large{(b)}}};
\end{axis}
\end{tikzpicture}
\\
\begin{tikzpicture}[baseline,trim axis left,scale=0.8]
\begin{axis}[
   trig format plots = rad,
   title={{\large{$\phi = \pi/4$}}},
   enlargelimits=false,
   thick,
   xmin=0, xmax=1,
    ymin=0, ymax=1.25,
   xlabel={$p$}
]
\addplot[
   red,
   very thick,
   domain=0:1,
   samples=300,
]
   {((19*x*x)/(4*(1+3*x)))};
   

    \addplot[
   blue,
   dashed,
   very thick,
   domain=0:1,
   samples=300,
]
   {(1/(2*(pi/6)*(pi/6)))*x*x*cos(pi/4)*cos(pi/4)*sin(pi/12)*sin(pi/12)*(9 - 3*cos(2*pi/4) + 8*cos(pi/6) + 4*cos(pi/4)*cos(pi/4)*cos(2*pi/6))};
   
   
           \addplot[
             black,
   very thick,
    dashdotted,
   domain=0:1,
   samples=300,
]
   {(1/((pi/6)*(pi/6)))*x*x*cos(pi/4)*cos(pi/4)*cos(pi/4)*cos(pi/4)*sin(pi/4)*sin(pi/4)};
   
   
      \addplot[
   gray,
   thick,
   dotted,
   domain=0:1,
   samples=300,
]
   {0.75};
   
   
           \addplot[
   brown,
   very thick,
   loosely dotted,
   domain=0:1,
   samples=300,
]
   {(19/16)*x*x};
   
   
              \addplot[
   magenta,
   very thick,
   loosely dashdotted,
   domain=0:1,
   samples=300,
]
   {(9/16)*x*x};
   
\node at (axis cs:0.1,1.1) {{\large{(c)}}};
\end{axis}
\end{tikzpicture}
&
\begin{tikzpicture}[baseline,trim axis right,scale=0.8]
\begin{axis}[
   trig format plots = rad,
   title={{\large{$\phi = \pi/3$}}},
   enlargelimits=false,
   thick,
   xmin=0, xmax=1,
    ymin=0, ymax=1,
   xlabel={$p$}
]
\addplot[
   red,
   very thick,
   domain=0:1,
   samples=300,
]
   {((39*x*x)/(16*(1+3*x)))};
   
   
    \addplot[
   blue,
   very thick,
   dashed,
   domain=0:1,
   samples=300,
]
   {(1/(2*(pi/6)*(pi/6)))*x*x*cos(pi/3)*cos(pi/3)*sin(pi/12)*sin(pi/12)*(9 - 3*cos(2*pi/3) + 8*cos(pi/6) + 4*cos(pi/3)*cos(pi/3)*cos(2*pi/6))};
   
   
           \addplot[
   black,
   very thick,
    dashdotted,
   domain=0:1,
   samples=300,
]
   {(1/((pi/6)*(pi/6)))*x*x*cos(pi/3)*cos(pi/3)*cos(pi/3)*cos(pi/3)*sin(pi/4)*sin(pi/4)};
   
   
        \addplot[
   gray,
   thick,
   dotted,
   domain=0:1,
   samples=300,
]
   {0.75};
   
           \addplot[
   brown,
   very thick,
   loosely dotted,
   domain=0:1,
   samples=300,
]
   {(39/64)*x*x};
   
   
              \addplot[
   magenta,
   very thick,
   loosely dashdotted,
   domain=0:1,
   samples=300,
]
   {(9/64)*x*x};
   
\node at (axis cs:0.1,0.88) {{\large{(d)}}};
\end{axis}
\end{tikzpicture}
    \\
\end{tabular}%
   \\
   \ref{named222}
\caption{(Color online) Example 3: Plot of quantum Fisher information ${F_Q}({\rho_{\tau}})$ (red solid line), squared speed ${\mathcal{S}_{\tau}}({\rho_0},H)$ (blue dashed line), ${\mathcal{B}_{{\tau},{m_{\text{max}}}}}({\rho_0},H)$ (black dot dashed line), ${F_I}({\rho_0},H)$ (brown loosely dotted line), and ${F_I^{m_{\text{max}}}}({\rho_0},H)$ (magenta loosely dot dashed line) related to the evolved state ${\rho_{\tau}} = {U_{\tau}}\,{\rho_0}{U_{\tau}^{\dagger}}$, where ${\rho_0} = \left(\frac{1 - p}{8}\right)\mathbb{I} + p\,|\Psi\rangle\langle{\Psi}|$ and $|\Psi\rangle = \cos{\phi}\,|{\text{GHZ}_3}\rangle + \sin{\phi}\,|{001}\rangle$ (${m_{\text{max}}}$), for (a)~$\phi = 0$, (b)~$\phi = \pi/6$, (c)~$\phi = \pi/4$, and (d)~$\phi = \pi/3$. Notice that the unitary evolution is generated by the Hamiltonian $H$ in Eq.~\eqref{eq:generatorNequal3} and $\tau = \pi/6$. The gray dotted line represents the QFI related to the three-qubits uniform superposition initial state ${\rho_0^{\text{class}}} = |{{+}{+}{+}}\rangle\langle{{+}{+}{+}}|$ which undergoes such unitary dynamics.}
\label{example3dot02}
\end{center}
\end{figure*}

Let us consider now the three-qubit initial mixed state
\begin{equation}
{\rho_0} = \left(\frac{1 - p}{8}\right)\mathbb{I} + p\,|\Psi\rangle\langle{\Psi}| ~,
\end{equation}
with 
\begin{equation}
|\Psi\rangle = \cos{\phi}\,|{\text{GHZ}_3}\rangle + \sin{\phi}\,|{{0}{0}{1}}\rangle ~,
\end{equation}
where the $|{\text{GHZ}_3}\rangle$ state is given in Eq.~\eqref{eq:ghzstate}, $0 \leq p \leq 1$, and $0 \leq \phi \leq \pi/2$. The probe state undergoes the unitary evolution ${\rho_{\tau}} = {U_{\tau}}\,{\rho_0}{U_{\tau}^{\dagger}}$ encoding on it the parameter $\tau$, where ${U_{\tau}} = {e^{-i\tau H}}$ and $H$ is given in Eq.~\eqref{eq:generatorNequal3}. After some calculations, the quantum Fisher information related to ${\rho_{\tau}}$ becomes
\begin{equation}
\label{eq:fisher_information_example3_3qubits}
{F_Q}({\rho_{\tau}}) = \frac{{p^2}(9 + {\cos^2}{\phi})\,{\cos^2}\phi}{2(1 + 3p)} ~.
\end{equation}
Opposite to the previous example, here $\rho_0$, ${\rho_{\tau}}$ and ${\rho^{\text{class}}_{\tau}}$ have ${m_{\text{max}}} = 3$ nonzero coherence order. In this case, both coherence quantifiers ${\mathcal{S}_{\tau}}({\rho_0},H)$ and ${\mathcal{B}_{{\tau},2}}({\rho_0},H)$ defined in Eq.~\eqref{danadona} can be written as follows
\begin{align}
\label{eq:fisher_information_example3_3qubits01}
{\mathcal{S}_{\tau}}({\rho_0},H) = \frac{p^2}{2{\tau^2}} &\left[9 - 3\cos(2\phi) + 8\cos\tau \right. \nonumber\\ 
& \left.+ 4\,{\cos^2}\phi\cos(2\tau)\right]\,{\cos^2}\phi\, {\sin^2}\left(\frac{\tau}{2}\right)
\end{align}
and
\begin{equation}
\label{eq:fisher_information_example3_3qubits02}
{\mathcal{B}_{{\tau},3}}({\rho_0},H) = \frac{1}{\tau^2}\,{\cos^4}\phi\,{\sin^2}\left(\frac{3\tau}{2}\right) ~.
\end{equation}
After some simpler calculations, we get
\begin{equation}
\label{eq:fisher_information_example3_3qubits02bbbbbb}
{F_I}({\rho_0},H) = \frac{1}{8}\,[19 - \cos(2\phi)]{p^2}{\cos^2}\phi
\end{equation}
and
\begin{equation}
\label{eq:fisher_information_example3_3qubits02bbbbbb02}
{F_I^{\,{m_{\text{max}}}}}({\rho_0},H) = \frac{9}{4}\,{p^2}{\cos^4}\phi ~.
\end{equation}
In Fig.~\ref{example3dot02} we plot the QFI related to the states ${\rho_{\tau}}$ and ${\rho_{\tau}^{\text{class}}}$, $\mathcal{S_{\tau}}({\rho_0},H)$, $\mathcal{B_{\tau}}({\rho_0},H)$, ${F_I}({\rho_0},H)$, and ${F_I^{\,{m_{\text{max}}}}}({\rho_0},H)$ given in Eqs.~\eqref{eq:fisher_information_example3_3qubits},~\eqref{eq:fisherexample10000b},~\eqref{eq:fisher_information_example3_3qubits01},
~\eqref{eq:fisher_information_example3_3qubits02},~\eqref{eq:fisher_information_example3_3qubits02bbbbbb}, and~\eqref{eq:fisher_information_example3_3qubits02bbbbbb02}, respectively.

\subsection{Example 4}

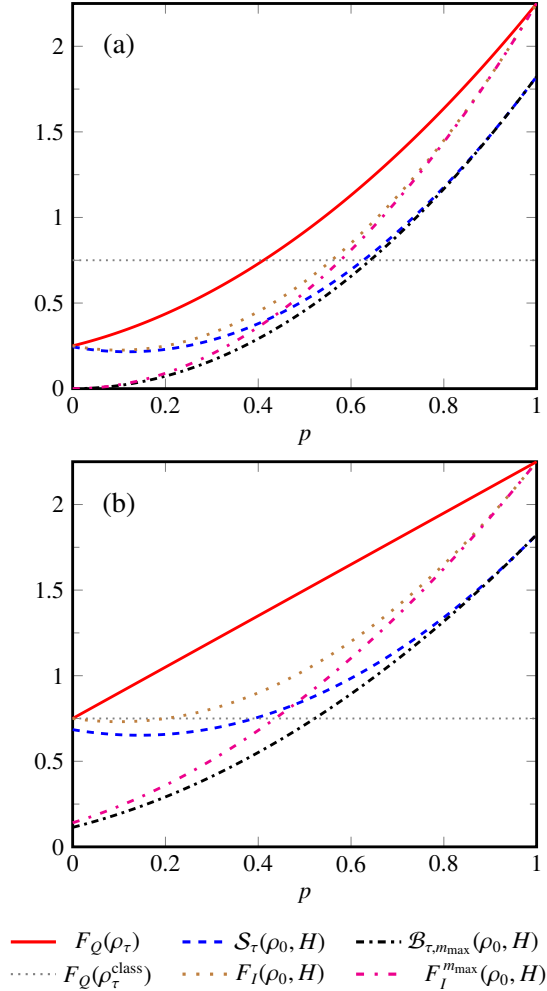
\begin{figure}[t]
\begin{tabular}{lr}
\begin{tikzpicture}[scale=0.9,font=\normalsize]
\begin{axis}[
  trig format plots = rad,
    enlargelimits=false,
    very thick,
    minor y tick num=1,
    legend style={draw=none},
    ymin=0, ymax=2.25,
    xmin=0, xmax=1,
    xlabel={$p$},
]
\addplot[
    red,
    very thick,
    domain=0:1,
    samples=100,
]
{(1/12)*(3 + 8*x*(1 + 2*x))};
\addplot[
   blue,
   very thick,
   dashed,
   domain=0:1,
   samples=100,
]
{(1/((pi/6)*(pi/6)))*sin(pi/12)*sin(pi/12)*(1 - 2*x + 4*x*x + 2*x*x*(2*cos(pi/6) + cos(2*pi/6)))};
     \addplot[
   black,
   very thick,
    dashdotted,
   domain=0:1,
   samples=300,
]
   {(1/((pi/6)*(pi/6)))*x*x*sin(pi/4)*sin(pi/4)};
\addplot[
   gray,
   thick,
   dotted,
   domain=0:1,
   samples=100,
]
{0.75};

   
           \addplot[
   brown,
   very thick,
   loosely dotted,
   domain=0:1,
   samples=300,
]
   {(1/4)*(1 - 2*x*(1 - 5*x))};
   
   
              \addplot[
   magenta,
   very thick,
   loosely dashdotted,
   domain=0:1,
   samples=300,
]
   {(9/4)*x*x};
\node at (axis cs:0.1,2) {{\large{(a)}}};
\end{axis}
\end{tikzpicture}
\\
\begin{tikzpicture}[scale=0.9,font=\normalsize]
\begin{axis}[
  trig format plots = rad,
    enlargelimits=false,
    very thick,
    minor y tick num=1,
    ymin=0, ymax=2.25,
    legend style={draw=none},
    xlabel={$p$},
    legend columns=3,
    legend entries={[black]${F_Q}({\rho_{\tau}})$~~~~~\\[black]${\mathcal{S}_{\tau}}({\rho_0},H)$~~~~~\\[black]${\mathcal{B}_{{\tau},{m_{\text{max}}}}}({\rho_0},H)$~~~~~\\[black]${F_Q}({\rho_{\tau}^{\text{class}}})$~~~~~\\[black]${F_I}({\rho_0},H)$~~~~~\\[black]${F_I^{\,{m_{\text{max}}}}}({\rho_0},H)$\\},
    legend to name=named2,
    legend style={draw=none,font=\small},
]
\addplot[
    red,
    very thick,
    domain=0:1,
    samples=100,
]
{(1/4)*(3 + 6*x)};
\addplot[
    blue,
    very thick,
    dashed,
    domain=0:1,
    samples=100,
]
{(1/(8*(pi/6)*(pi/6)))*sin(pi/12)*sin(pi/12)*( 3*(5 - 6*x + 9*x*x) + 8*cos(pi/6)*(1 + 3*x*x) + (1 + 3*x)*(1 + 3*x)*cos(2*pi/6))};
     \addplot[
   black,
   very thick,
    dashdotted,
   domain=0:1,
   samples=300,
]
   {(1/(16*(pi/6)*(pi/6)))*(1 + 3*x)*(1 + 3*x)*sin(pi/4)*sin(pi/4)};
\addplot[
   gray,
   thick,
   dotted,
   domain=0:1,
   samples=100,
]
{0.75};
   
           \addplot[
   brown,
   very thick,
   loosely dotted,
   domain=0:1,
   samples=300,
]
   {(3/8)*(2 - x*(1 - 5*x))};
   
   
              \addplot[
   magenta,
   very thick,
   loosely dashdotted,
   domain=0:1,
   samples=300,
]
   {(9/64)*(1 + 3*x)*(1 + 3*x)};
\node at (axis cs:0.1,2) {{\large{(b)}}};
\end{axis}
\end{tikzpicture}
\end{tabular}
\\   
\ref{named2}
\caption{(Color online) Plot of quantum Fisher information ${F_Q}({\rho_{\tau}})$ (red solid line), squared speed ${\mathcal{S}_{\tau}}({\rho_0},H)$ (blue dashed line), ${\mathcal{B}_{{\tau},{m_{\text{max}}}}}({\rho_0},H)$ (black dot dashed line), ${F_I}({\rho_0},H)$ (brown loosely dotted line), and ${F_I^{m_{\text{max}}}}({\rho_0},H)$ (magenta loosely dot dashed line) related to the evolved state ${\rho_{\tau}} = {U_{\tau}}\,{\rho_0}{U_{\tau}^{\dagger}}$, where (a)~example 4: ${\rho_0} = (1 - p)|{+00}\rangle\langle{+00}| + p|{\text{GHZ}_3}\rangle\langle{\text{GHZ}_3}|$ (${m_{\text{max}}} = 3$) and (b)~example 5: ${\rho_0} = (1 - p)|{{+}{+}{+}}\rangle\langle{{+}{+}{+}}| + p|{\text{GHZ}_3}\rangle\langle{\text{GHZ}_3}|$ (${m_{\text{max}}} = 3$). In both cases such unitary evolution is generated by the Hamiltonian $H$ in Eq.~\eqref{eq:generatorNequal3} and $\tau = \pi/6$. The gray dotted line represents the QFI related to the three-qubits uniform superposition initial state ${\rho_0^{\text{class}}} = |{+++}\rangle\langle{+++}|$, which undergoes the unitary dynamics discussed before.}
\label{example4and5dot0223}
\end{figure}

Let us consider the initial mixed state 
\begin{equation}
{\rho_0} = (1 - p)|{+00}\rangle\langle{+00}| + p|{\text{GHZ}_3}\rangle\langle{\text{GHZ}_3}| ~,
\end{equation}
with $0 \leq p \leq 1$ and the GHZ state given in Eq.~\eqref{eq:ghzstate}. The probe state $\rho_0$ undergoes the unitary evolution ${\rho_{\tau}} = {U_{\tau}}\,{\rho_0}{U_{\tau}^{\dagger}}$ which encodes the parameter $\tau$, where ${U_{\tau}} = {e^{-i\tau H}}$ and $H$ given in Eq.~\eqref{eq:generatorNequal3}. After some calculations the quantum Fisher information related to ${\rho_{\tau}}$ becomes
\begin{align}
\label{eq:fisherexample4}
{F_Q}({\rho_{\tau}}) = \frac{1}{12}\left[3 + 8p(1 + 2p)\right] ~.
\end{align}
Similarly to example 3, here the set of states $\rho_0$, ${\rho_{\tau}}$, and ${\rho^{\text{class}}_{\tau}}$ have ${m_{\text{max}}} = 3$ nonzero coherence order. Thus, it is straightforward to show that coherence quantifiers ${\mathcal{S}_{\tau}}({\rho_0},H)$ and ${\mathcal{B}_{{\tau},2}}({\rho_0},H)$ defined in Eq.~\eqref{danadona} read as
\begin{align}
\label{eq:fisherexample401}
\mathcal{S_{\tau}}({\rho_0},H) = \frac{1}{{\tau^2}} &\left[ 2{p^2}(\cos(2\tau) + 2\cos\tau) \right. \nonumber\\
&\left. 1 - 2p + 4{p^2} \right]\,{\sin^2}\left(\frac{\tau}{2}\right)
\end{align}
and
\begin{equation}
\label{eq:fisherexample402}
\mathcal{B_{\tau}}({\rho_0},H) = \frac{p^2}{\tau^2}\,{\sin^2}\left(\frac{3\tau}{2}\right) ~.
\end{equation}
Notice that one can verify that 
\begin{equation}
\label{eq:fisherexample402bbbbb}
{F_I}({\rho_0},H) = \frac{1}{4}\,[1 + 2p(5p - 1)]
\end{equation}
and 
\begin{equation}
\label{eq:fisherexample402bbbbb02}
{F_I^{m_{\text{max}}}}({\rho_0},H) = \frac{9}{4}\,{p^2} ~.
\end{equation}
In Fig.~\ref{example4and5dot0223}(a) we plot the quantum Fisher information with respect to the states ${\rho_{\tau}}$ and ${\rho_{\tau}^{\text{class}}}$, $\mathcal{S_{\tau}}({\rho_0},H)$, ${\mathcal{B}_{{\tau},3}}({\rho_0},H)$, ${F_I}({\rho_0},H)$, and ${F_I^{m_{\text{max}}}}({\rho_0},H)$ given in Eqs.~\eqref{eq:fisherexample4},~\eqref{eq:fisherexample10000b},~\eqref{eq:fisherexample401},~\eqref{eq:fisherexample402},~\eqref{eq:fisherexample402bbbbb}, and~\eqref{eq:fisherexample402bbbbb02}, respectively.

\subsection{Example 5}

Finally, let the initial mixed state be
\begin{equation}
{\rho_0} = (1 - p)|{{+}{+}{+}}\rangle\langle{{+}{+}{+}}| + p|{\text{GHZ}_3}\rangle\langle{\text{GHZ}_3}| ~,
\end{equation}
where $0 \leq p \leq 1$ and the GHZ state is given in Eq.~\eqref{eq:ghzstate}. Si\-mi\-lar\-ly to the previous examples, such an initial state undergoes the unitary evolution ${\rho_{\tau}} = {U_{\tau}}\,{\rho_0}{U_{\tau}^{\dagger}}$, where ${U_{\tau}} = {e^{-i\tau H}}$ and $H$ given in Eq.~\eqref{eq:generatorNequal3}. One can verify that the quantum Fisher information related to ${\rho_{\tau}}$ is given by
\begin{equation}
\label{eq:fisherexample5}
{F_Q}({\rho_{\tau}}) = \frac{3}{4}(1 + 2p) ~.
\end{equation}
We stress the fact that the states $\rho_0$, ${\rho_{\tau}}$, and ${\rho^{\text{class}}_{\tau}}$ have ${m_{\text{max}}} = 3$ nonzero coherence order and therefore the coherence quantifiers ${\mathcal{S}_{\tau}}({\rho_0},H)$ and ${\mathcal{B}_{{\tau},2}}({\rho_0},H)$ proposed in Eq.~\eqref{danadona} read as
\begin{align}
\label{eq:fisherexample501}
{\mathcal{S}_{\tau}}({\rho_0},H) = \frac{1}{8{\tau^2}} &\left[ 3(5 - 6p + 9{p^2}) + 8(1 + 3{p^2})\cos\tau \right. \nonumber \\ 
& \left. + {(1 + 3p)^2}\cos(2\tau)\right]\,{\sin^2}\left(\frac{\tau}{2}\right)
\end{align}
and
\begin{equation}
\label{eq:fisherexample502}
{\mathcal{B}_{{\tau},3}}({\rho_0},H) = \frac{{(1 + 3p)^2}}{16\,{\tau^2}}\,{\sin^2}\left(\frac{3\tau}{2}\right) ~.
\end{equation}
Moreover, the following can be shown: 
\begin{equation}
\label{eq:fisherexample502bbbbbb}
{F_I}({\rho_0},H) = \frac{3}{8}\,[2 + p(5p - 1)] ~,\quad {F_I^{m_{\text{max}}}}({\rho_0},H) = \frac{9}{64}{(1 + 3p)^2} ~.
\end{equation}
In Fig.~\ref{example4and5dot0223}(b) we plot the quantum Fisher information with respect to the states ${\rho_{\tau}}$ and ${\rho_{\tau}^{\text{class}}}$, $\mathcal{S_{\tau}}({\rho_0},H)$, ${\mathcal{B}_{{\tau},3}}({\rho_0},H)$, ${F_I}({\rho_0},H)$, and ${F_I^{m_{\text{max}}}}({\rho_0},H)$ given in Eqs.~\eqref{eq:fisherexample5},~\eqref{eq:fisherexample10000b},~\eqref{eq:fisherexample501},~\eqref{eq:fisherexample502}, and~\eqref{eq:fisherexample502bbbbbb}, respectively.


%

\end{document}